\documentclass[11pt]{article}

\usepackage{graphicx,color}
\usepackage{amsmath,amssymb}

\begin{document}

\title{On the Coupling-Strength Growth \\ of the Rabi Model \\ 
in the Light of SUSYQM}

\author{Masao Hirokawa\thanks{Department of Mathematics, Okayama University, 
Okayama, 700-8530, Japan ({\tt hirokawa@math.okayama-u.ac.jp}).}}

\date{\today}

\markright{M. Hirokawa}
\pagestyle{myheadings}

\maketitle

\begin{abstract}
We consider the coupling-strength growth of the Rabi model 
from the point of the view of SUSYQM. 
We show that the Rabi model takes the supersymmetric 
system to the spontaneous supersymmetry breaking 
as its coupling strength $\mathrm{g}$ grows lager 
from the case $\mathrm{g}=0$ to the case $\mathrm{g}\approx\infty$.  
We study a kind of chirality quantum phase transition 
in this process. 
\end{abstract}

\section{Introduction}
\label{sec:intro}

Supersymmetric quantum mechanics (SUSYQM) was initiated 
by Witten \cite{Witten}, and has been developed 
by many physicists 
\cite{SH82,CR83,CH85,DHV85,balantekin,DUZ85,JLL87,Arai89,SS08,GMR,Schwabl,Sugino}. 
In particular, some ground state structures and the spontaneous 
supersymmetry (SUSY) breaking 
in SUSYQM have been investigated \cite{CH85,JLL87,Arai89,Sugino}. 
We are interested in ground state structure 
in SUSYQM from another point of view than 
their preceding studies. 
We will handle the Rabi model that has the interaction 
between a $2$-level atom and the light in a cavity. 
The Rabi model is sometimes called the full Jaynes-Cummings model. 
That is, its Hamiltonian has full linear coupling 
of the $2$-level atom and the light 
without the rotating wave approximation (RWA). 
The Rabi model has been well studied 
in quantum optics, 
and its some inherent properties 
have been beginning to experimentally observed 
in cavity quantum electrodynamics (QED) and circuit QED 
\cite{HR,RBH,MB-MSS,BHGS,Chiorescu-hofheinz09,Wallraff08,Gross-Mooij}. 
We take an interest in the physical properties 
that a qubit of the $2$-level atom coupled with the light makes in SUSYQM. 
We are conjecturing that the spontaneous SUSY breaking recovers 
a chirality in the Rabi model. 

The interaction between an atom and the light 
in nature follows the QED. 
It is governed by the fine-structure constant 
$\alpha\approx 0.00729735$, 
belonging to the region over which 
the perturbation theory is valid.  
On the other hand, cavity QED handles 
stronger interaction than the standard QED does \cite{HR,RBH}.  
Such a strong interaction is experimentally 
prepared with the coupling of 
a two-level atom and a one-mode light (i.e., single-mode laser) 
in a mirror cavity (i.e., a mirror resonator). 
Several solid-state analogues of the strong coupling 
had been foreseen in superconducting systems
\cite{MB-MSS,BHGS}. 
In short, we respectively replace 
the atom, the light, and the mirror resonator in cavity QED 
by an artificial atom, 
a microwave, and a microwave resonator 
on a superconducting circuit. 
The artificial atom consists of 
a superconducting circuit based on some Josephson junctions then. 
This replaced cavity QED is circuit QED, 
which has been experimentally demonstrated 
\cite{Chiorescu-hofheinz09,Wallraff08}. 
It is remarkable that circuit QED has been capable 
of intensifying the coupling strength 
further than cavity QED has \cite{Gross-Mooij}.

In this paper we pay our particular attention on 
the coupling-strength growth of the Rabi model 
from the point of the view of SUSYQM. 
We will show that the Rabi model takes 
SUSY system to the spontaneous SUSY breaking 
as its coupling strength $\mathrm{g}$ grows lager 
from the case $\mathrm{g}=0$ to the case $\mathrm{g}\approx\infty$. 
We will also show that this spontaneous SUSY breaking is caused by 
the spin-chirality between the two levels of the atom. 
We are then interested in when and how the spin-chirality works. 
We will consider a problem similar to Hund's paradox 
on the chiral molecules \cite{hund, wightman}, 
and show a kind of chirality quantum phase transition 
(CQPT) \cite{BM07-08} in the process 
from the SUSY system to the system with spontaneous SUSY breaking. 
This makes us expect that we can realize 
the quantum simulation \cite{feynman82} 
of some properties of SUSYQM in circuit QED.

\section{SUSY and Spontaneous Breaking in Rabi Model}
\label{sec:SUSY-Breaking} 

In this section we consider two special cases 
to find some SUSYQM aspects in the Rabi model.

We denote the annihilation (resp. creation) operator 
for the one-mode photon by $a$ (resp. $a^{\dagger}$). 
We use the standard notation for the Pauli matrices 
as  
$\sigma_{x}
\equiv 
\bigl(\begin{smallmatrix}
0 & 1 \\ 
1 & 0 
\end{smallmatrix}\bigr)$, 
$\sigma_{y}\equiv 
\bigl(\begin{smallmatrix}
0 & -i \\ 
i & 0 
\end{smallmatrix}\bigr)$, 
and 
$\sigma_{z}\equiv 
\bigl(\begin{smallmatrix}
1 & 0 \\ 
0 & -1 
\end{smallmatrix}\bigr)$. 
We define spin states $|\!\!\uparrow\rangle$ 
and $|\!\!\downarrow\rangle$ by 
$|\!\!\uparrow\rangle :=\bigl(
\begin{smallmatrix}
1 \\ 0 
\end{smallmatrix}
\bigr)$ and
$|\!\!\downarrow\rangle :=\bigl( 
\begin{smallmatrix}
0 \\ 1 
\end{smallmatrix}
\bigr)$. 
We denote by $\mathcal{F}$ the Fock space 
of the one-mode photon, 
and by $|n\rangle$ the Fock state with 
the photon number $n=0, 1, 2, \cdots$. 
So, in particular, $|0\rangle$ denotes 
the Fock vacuum. 
Every quantum state that we will use in this paper 
is represented as $|n,\sharp\rangle:=
|n\rangle|\sharp\rangle$ for 
$n=0, 1, 2, \cdots$ and $\sharp=\uparrow, \downarrow$. 
Here, we omitted the tensor-notation $\otimes$ 
from the expression $|n\rangle\otimes|\sharp\rangle$. 
We will use this omitted notation throughout this paper.  
Also, we denote by $|\psi,\sharp\rangle$ 
the state $|\psi\rangle|\sharp\rangle$ 
for the state $\psi$ in the Fock space $\mathcal{F}$.  
Let us give the subspace $\mathcal{H}_{\mathrm{even}}$ 
(resp. $\mathcal{H}_{\mathrm{odd}}$) as 
the set of all of the states 
$|\psi, \uparrow\rangle$ 
(resp. $|\psi, \downarrow\rangle$), 
where the state $\psi$ runs over the whole Fock space. 
Then, the state space $\mathcal{H}$ is obviously 
decomposed as the direct sum of $\mathcal{H}_{\mathrm{even}}$ 
and $\mathcal{H}_{\mathrm{odd}}$: 
$\mathcal{H}=\mathcal{H}_{\mathrm{even}}\oplus\mathcal{H}_{\mathrm{odd}}$.

\begin{figure}[htbp]
  \begin{center}
  \resizebox{70mm}{!}{\includegraphics{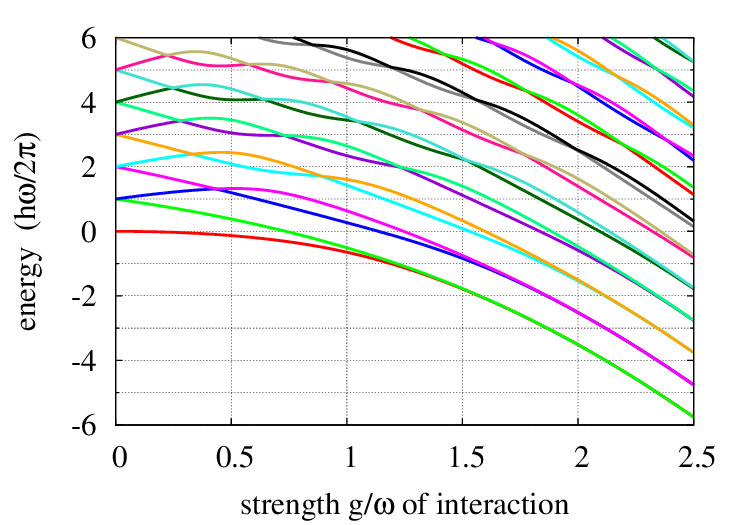}} 
  \end{center}
  \vspace*{2mm}
  \caption{\scriptsize 
Energy Levels of $H_{\mathrm{Rabi}}$ 
for $\omega:=
\omega_{\mathrm{a}}=\omega_{\mathrm{c}}$. 
Each color indicates the $n$th level of the energy 
for $n=0, 1, 2, \cdots$ from the bottom, 
where the $0$th level energy means the ground-state energy.}
\label{fig:Rabi}
\end{figure}

The free Hamiltonian $H_{0}$ 
of the Rabi model is given by 
$$
H_{0}:=
\frac{\hbar\omega_{\mathrm{a}}}{2}\sigma_{z}
+\hbar\omega_{\mathrm{c}}
\left(a^{\dagger}a+\frac{1}{2}\right). 
$$ 
The constants $\omega_{\mathrm{a}}$ 
and $\omega_{\mathrm{c}}$ are respectively 
the (artificial) atom transition frequency and 
the cavity resonance frequency.  
Then, the Rabi Hamiltonian $H_{\mathrm{Rabi}}$ is given by 
\begin{equation}
H_{\mathrm{Rabi}}:=
H_{0}
+\hbar\mathrm{g}\left( a+a^{\dagger}\right)\sigma_{x},
\label{eq:Rabi-Hamiltonian-0}
\end{equation} 
where the parameter $\mathrm{g}\ge 0$ stands for 
the atom-photon coupling constant 
that represents the coupling strength. 
The solvability of the Rabi Hamiltonian 
has been argued by Braak \cite{Braak11}, 
by using Bargmann's representation \cite{RD}. 
Here, we give a numerical computation 
of the energies of the Rabi model 
in the case $\omega:=\omega_{\mathrm{a}}=
\omega_{\mathrm{c}}$ in Fig.\ref{fig:Rabi}.  

Fig.\ref{fig:Rabi} attracts our particular attention to 
the two special cases, $\mathrm{g}=0$ and $\mathrm{g}\approx\infty$, 
in the light of SUSYQM. 
In the case $\mathrm{g}=0$, 
the ground state is unique, but all the excited states 
are $2$-fold degenerate. 
All the eigenenergies line up at an equal interval $\hbar\omega$ then.  
Meanwhile, in case $\mathrm{g}\approx\infty$, 
Fig.\ref{fig:Rabi} makes us expect that  
all the states are almost $2$-fold degenerate, 
and all the eigenenergies are aligned at an almost equal interval $\hbar\omega$. 
We now investigate these physical situations in detail 
from the point of the view of SUSYQM for a while. 
Let us take the two frequencies as $\omega:=
\omega_{\mathrm{a}}=\omega_{\mathrm{c}}$ 
throughout this paper, 
and then, we denote the free Hamiltonian $H_{0}$ 
by $H_{\mathrm{SS}}$: 
\begin{equation}
H_{\mathrm{SS}}:=
\frac{\hbar\omega}{2}\sigma_{z}
+\hbar\omega\left(a^{\dagger}a+\frac{1}{2}\right). 
\label{eq:HSS}
\end{equation}
We note that it is easy to control the two 
frequencies $\omega_{\mathrm{a}}$ and $\omega_{\mathrm{c}}$ 
in circuit QED so that they are equal.  

First up, when there is no interaction between 
the (artificial) atom and the light (i.e., $\mathrm{g}=0$), 
the Rabi Hamiltonian becomes the free Hamiltonian: 
$H_{\mathrm{Rabi}}=H_{\mathrm{SS}}$. 
So, it is the most popular Hamiltonian in SUSYQM: 
$H_{\mathrm{Rabi}}=(1/2m)\left(
p^{2}+W^{2}+\hbar(dW/dx)\sigma_{z}
\right)$ 
with the correspondence, 
$a=\sqrt{m\omega/2\hbar}\, x+ip/\sqrt{2m\hbar\omega}$ 
and 
$a^{\dagger}=\sqrt{m\omega/2\hbar}\, x-ip/\sqrt{2m\hbar\omega}$ 
for the position operator $x$ and the momentum operator $p$, 
where the superpotential $W$ is given by 
$W(x)=m\omega x$. 
Namely, the system has $N=2$\, SUSY. 
More precisely, 
the supercharges $Q_{1}$ and $Q_{2}$ defined by 
$Q_{1}=(1/2\sqrt{m})(W\sigma_{x}-p\sigma_{y})$ and 
$Q_{2}=(1/2\sqrt{m})(W\sigma_{y}+p\sigma_{x})$ 
make the relations: 
\begin{align*}
& \left\{ Q_{k} , Q_{\ell}\right\}=\delta_{k\ell}H_{\mathrm{Rabi}}, \\ 
& \left[ Q_{k} , H_{\mathrm{Rabi}}\right]=0, \\ 
& \left\{ Q_{k} , N_{\mathrm{F}}\right\}=0,  
\end{align*} 
for $k, \ell=1, 2$, 
and the grading operator $N_{\mathrm{F}}=\sigma_{z}$ 
satisfying the conditions, 
$N_{\mathrm{F}}\psi_{\mathrm{even}}=\psi_{\mathrm{even}}$ 
for any state $\psi_{\mathrm{even}}\in\mathcal{H}_{\mathrm{even}}$, 
and $N_{\mathrm{F}}\psi_{\mathrm{odd}}=\, -\psi_{\mathrm{odd}}$ 
for any state $\psi_{\mathrm{odd}}\in\mathcal{H}_{\mathrm{odd}}$. 
Here the symbol $\delta_{k\ell}$ is the Kronecker delta. 
Then, the system has no SUSY breaking. 
That is, the supersymmetric (SUSY) ground state is 
$|0, \downarrow\rangle$ and therefore 
the ground state energy is equal to zero in this case. 
In addition,  for $\mathrm{g}=0$, 
the ground state of the Rabi Hamiltonian is unique \cite{HH12}, 
but all of its excited states are $2$-fold degenerate. 
The two degenerate excited states are interchanged with each other 
by the SUSY-generating charges $Q^{+}$ and $Q^{-}$ 
defined by 
$Q^{+}:=\sqrt{\hbar\omega}\, a\sigma_{+}$ and 
$Q^{-}:=\sqrt{\hbar\omega}\, a^{\dagger}\sigma_{-}$. 
Here, $\sigma_{-}$ and $\sigma_{+}$ are the spin 
annihilation and creation operators defined 
by $\sigma_{\pm}:=(\sigma_{x}\pm i\sigma_{y})/2$: 
\begin{align*}
& Q^{-}|n, \downarrow\rangle =
Q^{+}|n, \uparrow\rangle = 0, \\ 
& |n, \uparrow\rangle
=\frac{1}{\sqrt{\hbar\omega(n+1)}}
Q^{+}|n+1, \downarrow\rangle, \\ 
& |n+1, \downarrow\rangle
=\frac{1}{\sqrt{\hbar\omega(n+1)}}
Q^{-}|n, \uparrow\rangle,  
\end{align*}
for $n= 0, 1, 2, \cdots$. 
As is well known, of course, we have 
$$
\left\{ Q^{\pm} , Q^{\pm}\right\} = 0\,\,\,\text{and}\,\,\, 
H_{\mathrm{Rabi}}=\left\{ Q^{+} , Q^{-}\right\}.  
$$

On the other hand, let us take the coupling strength $\mathrm{g}$ 
large enough now. 
We can easily expect that the photon part energy, 
$$
H_{\mathrm{asym}}:=\hbar\omega\left( a^{\dagger}a+\frac{1}{2}\right) 
+\hbar\mathrm{g}\left( a+a^{\dagger}\right)\sigma_{x},
$$
is asymptotically much more dominant than 
the $2$-level atom energy $\hbar\omega\sigma_{z}/2$. 
Namely, the atom energy $\hbar\omega\sigma_{z}/2$ 
works as a very small perturbation for the photon part energy 
$H_{\mathrm{asym}}$ around $\mathrm{g}\approx\infty$ 
\cite{kus}.   
We can show this in a mathematically exact way. 
We define a unitary operator $U_{\mathrm{g}}$ by 
$$
U_{\mathrm{g}}:=\frac{1}{\sqrt{2}}
\begin{pmatrix}
V_{-} & - V_{+} \\ 
V_{-} & V_{+}  
\end{pmatrix}
$$
with the unitary operator $V_{\pm}:=e^{\pm\mathrm{g}(a^{\dagger}-a)/\omega}$. 
Recall the well-known Bogoliubov transformation: 
\begin{equation}
V_{\pm}\left\{ 
\hbar\omega 
\left( a^{\dagger}a+\frac{1}{2}\right) 
\pm\hbar\mathrm{g}
\left( a+a^{\dagger}\right)
\right\} 
V_{\mp}  
= 
\hbar\omega 
\left( a^{\dagger}a+\frac{1}{2}\right) 
-\hbar\frac{\mathrm{g}^{2}}{\omega}.  
\label{eq:bogoliubov}
\end{equation}
We reach the unitary transformation, 
\begin{equation}
U_{\mathrm{g}}^{*}H_{\mathrm{Rabi}}U_{\mathrm{g}}
=U_{\mathrm{g}}^{*}H_{\mathrm{asym}}U_{\mathrm{g}}
-\, \frac{\hbar\omega}{2}\widetilde{V}_{\mathrm{g}}
=\widetilde{H}_{0}
-\, \hbar\frac{\mathrm{g}^{2}}{\omega}
-\, \frac{\hbar\omega}{2}\widetilde{V}_{\mathrm{g}}
\label{eq:new-*1}
\end{equation} 
with the asymptotically free Hamiltonian
$$
\widetilde{H}_{0}=
\begin{pmatrix}
\hbar\omega \left( a^{\dagger}a+\frac{1}{2}\right) & 0 \\ 
0 & \hbar\omega \left( a^{\dagger}a+\frac{1}{2}\right)
\end{pmatrix}
$$
and the unitary, self-adjoint interaction
$$
\widetilde{V}_{\mathrm{g}}
=
\begin{pmatrix}
0 & e^{2\mathrm{g}(a^{\dagger}-a)/\omega} \\ 
e^{-2\mathrm{g}(a^{\dagger}-a)/\omega} & 0 \\ 
\end{pmatrix}. 
$$
For arbitrary wave functions 
$\psi=\bigl(\begin{smallmatrix}\psi_{1} \\ 
\psi_{2}\end{smallmatrix}\bigr)$ 
and $\phi=\bigl(\begin{smallmatrix}\phi_{1} \\ 
\phi_{2}\end{smallmatrix}\bigr)$, 
we set $\widetilde{\psi}_{j}:= 
e^{i\pi a^{\dagger}a/2}\psi_{j}$ and 
$\widetilde{\phi}_{j}:= 
e^{i\pi a^{\dagger}a/2}\phi_{j}$. 
Using the equations, 
$e^{i\pi a^{\dagger}a/2}ae^{-i\pi a^{\dagger}a/2}=-ia$ 
and $e^{i\pi a^{\dagger}a/2}a^{\dagger}e^{-i\pi a^{\dagger}a/2}=ia$, 
we have 
\begin{align*}
\langle\psi|\widetilde{V}_{\mathrm{g}}|
\phi\rangle 
&= 
\langle\psi_{1}|e^{2\mathrm{g}(a^{\dagger}-a)/\omega}\phi_{2}\rangle
+\langle\psi_{2}|e^{-2\mathrm{g}(a^{\dagger}-a)/\omega}\phi_{1}\rangle \\ 
&= \langle\widetilde{\psi}_{1}|
e^{i(2\mathrm{g}/\omega)(a^{\dagger}+a)}\widetilde{\phi}_{2}\rangle 
+\langle\widetilde{\psi}_{2}|
e^{-i(2\mathrm{g}/\omega)(a^{\dagger}+a)}\widetilde{\phi}_{1}\rangle. 
\end{align*}
Since we have 
$a^{\dagger}+a=\sqrt{2m\omega/\hbar}\, x$, 
we obtain the representation: 
$$
\langle\psi|\widetilde{V}_{\mathrm{g}}|
\phi\rangle 
= 
\int dx\widetilde{\psi}_{1}^{*}(x)\widetilde{\phi}_{2}(x)
e^{i(2\mathrm{g}\sqrt{2m/\hbar\omega})x}  
+ 
\int dx\widetilde{\psi}_{2}^{*}(x)\widetilde{\phi}_{1}(x)
e^{-i(2\mathrm{g}\sqrt{2m/\hbar\omega})x}. 
$$
The Riemann-Lebesgue's theorem tells us 
the term vanishes as 
\begin{equation}
\lim_{\mathrm{g}\to\infty}
\langle\psi|\widetilde{V}_{\mathrm{g}}|
\phi\rangle 
= 0  
\label{eq:weak-decay}
\end{equation}
in spite of the equation,
\begin{equation}
\langle\psi|\widetilde{V}_{\mathrm{g}}^{*}\widetilde{V}_{\mathrm{g}}|
\psi\rangle=\langle\psi|\psi\rangle.   
\label{eq:non-decay}
\end{equation} 
The weak decay (\ref{eq:weak-decay}) supplies us 
with the weak convergence of 
the operator 
$U_{\mathrm{g}}^{*}(H_{\mathrm{Rabi}}+\hbar\mathrm{g}^{2}/\omega)U_{\mathrm{g}}$: 
\begin{align}
\lim_{\mathrm{g}\to\infty}
\langle\psi|
U_{\mathrm{g}}^{*}(H_{\mathrm{Rabi}}+\hbar\mathrm{g}^{2}/\omega)U_{\mathrm{g}}
|\phi\rangle 
=&
\langle\psi|
U_{\mathrm{g}}^{*}(H_{\mathrm{asym}}+\hbar\mathrm{g}^{2}/\omega)U_{\mathrm{g}}
|\phi\rangle 
\notag \\ 
=&
\langle\psi|
\widetilde{H}_{0}
|\phi\rangle 
\label{eq:weak-convergence}
\end{align}
for arbitrary wave functions $\psi$ and $\phi$. 
We here point out that the vector 
$U_{\mathrm{g}}^{*}(H_{\mathrm{Rabi}}+\hbar\mathrm{g}^{2}/\omega)U_{\mathrm{g}}
|\phi\rangle$ never converges to the vector 
$\widetilde{H}_{0}|\phi\rangle$ 
in the sense of the norm induced 
by the inner product of the Hilbert space 
$\mathcal{F}\otimes\mathbb{C}^{2}$. 
Otherwise, we have the limit, 
$\lim_{\mathrm{g}\to\infty}
\widetilde{V}_{\mathrm{g}}|\phi\rangle=0$, 
in the norm sense. 
It, however, contradicts Eq.(\ref{eq:non-decay}).  

As explained precisely in \S\ref{sec:justification}, 
the weak convergence (\ref{eq:weak-convergence}) 
makes a correspondence 
between eigenstates $\varphi^{\mathrm{Rabi}}$ of 
the Rabi Hamiltonian $H_{\mathrm{Rabi}}$ 
and eigenstates $\varphi_{n}$ 
of the asymptotic Hamiltonian 
$H_{\mathrm{asym}}$ in the following: 
\begin{equation}
\varphi^{\mathrm{Rabi}}
\approx 
\varphi_{n},\qquad 
\mathrm{g}\gg 1, 
\label{eq:asymp-behavior}
\end{equation}
where $n$ is the non-negative integer satisfying 
the condition for the eigenenergy 
$E^{\mathrm{Rabi}}$ of the eigenstate 
$\varphi^{\mathrm{Rabi}}$: 
\begin{equation}
E^{\mathrm{Rabi}}
\approx 
\hbar\omega\left( n+\frac{1}{2}\right)
-\hbar\frac{\mathrm{g}^{2}}{\omega},\qquad 
\mathrm{g}\gg 1. 
\label{eq:asymp-behavior'}
\end{equation} 
Actually, we can chose the eigenstate 
$\varphi_{n}$ as either of one of eigenvectors 
$U_{\mathrm{g}}|n,\uparrow\rangle$ 
and $U_{\mathrm{g}}|n,\downarrow\rangle$. 
Eqs.(\ref{eq:asymp-behavior}) and 
(\ref{eq:asymp-behavior'}) show 
asymptotically $2$-fold degenerate 
energy levels as in Fig.\ref{fig:Rabi}. 
In particular, Eq.(\ref{eq:asymp-behavior}) says that 
the ground-state energy $E_{\mathrm{Rabi}}$ 
of the Rabi Hamiltonian has the asymptotic 
behavior as $E_{\mathrm{Rabi}}\approx 
\hbar\omega/2-\hbar\mathrm{g}^{2}/\omega$. 
This is justified with another method. 
See expressions in Eqs.(\ref{eq:gse-sb1}) and (\ref{eq:gse-sb2}) 
for more precise expression obtained by 
using the structure \`{a} la instanton gas.   

We here recall that the Rabi Hamiltonian has the 
following parity symmetry: 
$\left[ H_{\mathrm{Rabi}} , \Pi\right]=0$ 
for the parity operator $\Pi:=\sigma_{z}(-1)^{a^{\dagger}a}$. 
So, adopting the representations 
$b:=\sigma_{x}a$ and $b^{\dagger}:=\sigma_{x}a^{\dagger}$ 
satisfying the canonical commutation relation, 
$[b , b^{\dagger}]=1$, 
the Rabi Hamiltonian has the expression 
$H_{\mathrm{Rabi}}
=H_{\mathrm{asym}}+\hbar\omega\sigma_{z}/2$ with 
\begin{equation}
H_{\mathrm{asym}}
=\hbar\omega
\left( b^{\dagger}b+\frac{1}{2}\right)
+\hbar\mathrm{g}
\left( b^{\dagger}+b\right)
\label{eq:parity}
\end{equation}
and 
$$ 
\frac{\hbar\omega}{2}\sigma_{z}
=\frac{\hbar\omega}{2}(-1)^{b^{\dagger}b}\Pi 
\longrightarrow 0\quad 
\text{as}\,\,\, \mathrm{g}\to\infty 
$$
in the weak sense by Eq.(\ref{eq:weak-convergence}). 
Eq.(\ref{eq:parity}) tells us the energy that 
we have to renormalize. 
Thus, based on this expression and Eq.(\ref{eq:asymp-behavior}), 
we define the asymptotically renormalized (AR) Rabi Hamiltonian 
$H_{\mathrm{Rabi}}^{\mathrm{AR}}$ as: 
$$
H_{\mathrm{Rabi}}^{\mathrm{AR}}:=
H_{\mathrm{asym}}+\hbar\frac{\mathrm{g}^{2}}{\omega}. 
$$
We define the system's supercharges 
$Q_{1}$ and $Q_{2}$ by $Q_{1}:=U_{\mathrm{g}}\widetilde{Q}_{1}U_{\mathrm{g}}^{*}$ 
and $Q_{2}:=:=U_{\mathrm{g}}\widetilde{Q}_{2}U_{\mathrm{g}}^{*}$, 
where the operators $\widetilde{Q}_{1}$ and $\widetilde{Q}_{2}$ are given by 
\begin{align*}
& \widetilde{Q}_{1}:=\sqrt{\frac{\hbar\omega}{2}}\, 
\sqrt{a^{\dagger}a+\frac{1}{2}\,}\,\,\sigma_{x}, \\ 
& \widetilde{Q}_{2}:=\sqrt{\frac{\hbar\omega}{2}}\, 
\left(-i\sqrt{a^{\dagger}a+\frac{1}{2}\,}\,\,\sigma_{+}
+i\sqrt{a^{\dagger}a+\frac{1}{2}\,}\,\,\sigma_{-}\right),
\end{align*} 
in the present case. 
Then, we have the relations, 
\begin{align*}
& \left\{ Q_{k} , Q_{\ell}\right\}
=\delta_{k\ell}H_{\mathrm{Rabi}}^{\mathrm{AR}}, \\ 
& [ Q_{k} , H_{\mathrm{Rabi}}^{\mathrm{AR}}]=0, \\ 
& \left\{ Q_{k} , N_{\mathrm{F}}\right\}=0,  
\end{align*} 
for $k, \ell=1, 2$, 
and the grading operator $N_{\mathrm{F}}=U_{\mathrm{g}}\sigma_{z}U_{\mathrm{g}}^{*}=\sigma_{x}$, 
which satisfies the conditions, 
$N_{\mathrm{F}}\psi_{\mathrm{even}}=\psi_{\mathrm{even}}$ 
for any state $\psi_{\mathrm{even}}\in U_{\mathrm{g}}\mathcal{H}_{\mathrm{even}}$, 
and $N_{\mathrm{F}}\psi_{\mathrm{odd}}=\, -\psi_{\mathrm{odd}}$ 
for any state $\psi_{\mathrm{odd}}\in U_{\mathrm{g}}\mathcal{H}_{\mathrm{odd}}$. 
We note the equation concerning the whole state space, 
$\mathcal{H}=U_{\mathrm{g}}\mathcal{H}=(U_{\mathrm{g}}\mathcal{H}_{\mathrm{even}})
\oplus (U_{\mathrm{g}}\mathcal{H}_{\mathrm{odd}})$.  
We have the SUSY-generating supercharge 
$Q^{-}$ and $Q^{+}$ as 
$Q^{-}:=U_{\mathrm{g}}\widetilde{Q}^{-}U_{\mathrm{g}}^{*}$ and 
$Q^{+}:=U_{\mathrm{g}}\widetilde{Q}^{+}U_{\mathrm{g}}^{*}$, 
where the operators $\widetilde{Q}^{-}$ and $\widetilde{Q}^{+}$ are given by  
$$
\widetilde{Q}^{-}=\sqrt{\hbar\omega
\left( a^{\dagger}a+\frac{1}{2}\right)}\,\,\sigma_{+}\,\,\,\text{and}\,\,\, 
\widetilde{Q}^{+}=\sqrt{\hbar\omega
\left( a^{\dagger}a+\frac{1}{2}\right)}\,\,\sigma_{-}
$$
satisfying the relations:
$$
\left\{ Q^{\pm} , Q^{\pm}\right\} = 0\,\,\,\text{and}\,\,\, 
H_{\mathrm{Rabi}}^{\mathrm{AR}}=\left\{ Q^{+} , Q^{-}\right\}, 
$$
and moreover, 
\begin{align*}
& Q^{-}U_{\mathrm{g}}|n, \uparrow\rangle =
Q^{+}U_{\mathrm{g}}|n, \downarrow\rangle = 0, \\ 
& U_{\mathrm{g}}|n, \uparrow\rangle
=\frac{1}{\sqrt{\hbar\omega(n+1/2)}}
Q^{+}U_{\mathrm{g}}|n, \uparrow\rangle, \\ 
& U_{\mathrm{g}}|n, \downarrow\rangle
=\frac{1}{\sqrt{\hbar\omega(n+1/2)}}
Q^{-}U_{\mathrm{g}}|n, \downarrow\rangle,  
\end{align*}
for $n= 0, 1, 2, \cdots$. 

Using the equations, 
$[\sigma_{z} , \widetilde{H}_{0}]=0$ 
and $U_{\mathrm{g}}\sigma_{z}U_{\mathrm{g}}^{*}=\sigma_{x}$, 
we have the following symmetry:
\begin{equation}
[\sigma_{x} , H_{\mathrm{Rabi}}^{\mathrm{AR}}]=0. 
\label{eq:spin-chiral-symmetry}
\end{equation}
Define the states $\psi_{+}$ and $\psi_{-}$ by 
$\psi_{\pm}:=U_{\mathrm{g}}(|0,\uparrow\rangle\pm |0,\downarrow\rangle)/\sqrt{2}$. 
Then, the states  $\psi_{+}$ and $\psi_{-}$ are 
the lowest-energy states of the AR Rabi Hamiltonian 
$H_{\mathrm{Rabi}}^{\mathrm{AR}}$ 
since the states $(|0,\uparrow\rangle\pm 
|0,\downarrow\rangle)/\sqrt{2}$ 
are the lowest-energy states of the asymptotically free Hamiltonian 
$\widetilde{H}_{0}$.  
We then reach the fact that 
\begin{equation}
\sigma_{x}\psi_{+}=\psi_{-}\ne\psi_{+}\,\,\,\text{with}\,\,\, 
\langle\psi_{+}|\psi_{-}\rangle=0.
\label{eq:spin-chiral-relation}
\end{equation}
We here remember that the Pauli matrix $\sigma_{x}$ makes 
the spin-chiral transformation: 
$\sigma_{x}|\!\!\uparrow\rangle
=|\!\!\downarrow\rangle$ 
and 
$\sigma_{x}|\!\!\downarrow\rangle
=|\!\!\uparrow\rangle$. 
Therefore, \textit{although the AR Rabi Hamiltonian 
$H_{\mathrm{Rabi}}^{\mathrm{AR}}$ has the spin-chiral symmetry 
(\ref{eq:spin-chiral-symmetry}), 
the lowest-energy state is not invariant under 
the spin-chirality as in the relation 
(\ref{eq:spin-chiral-relation})}. 
This is exactly the spontaneous SUSY breaking 
that we are interested in. 
Thus, the system does not have the SUSY ground state, i.e., 
the lowest energy of the AR Rabi Hamiltonian is $\hbar\omega/2$ 
and it is strictly positive. 
Then, all of the energy states of the AR Rabi Hamiltonian 
are $2$-fold degenerate.

From these arguments we eventually realize that 
the growth of the coupling strength of the Rabi model 
plays a role of taking the $N=2$\, SUSY 
to the spontaneous SUSY breaking. 
We are interested in the process of 
the coupling strength's growth, 
which breaks the SUSY. 
Therefore, from the next section 
we will study when and how the effect of 
the spin-chirality appears in the Rabi Hamiltonian. 

We note here that in Ref.\cite{SM91} Schmitt and Mufti stated 
that they found a SUSY in the Rabi model for a case 
employing the two approximations. 
However, it has not been shown the proof without 
the approximations.

\section{Equitableness of Spin-Chirality in Rabi Model} 

According to several experimental results 
in cavity QED or circuit QED, 
the spin-chirality in the Rabi model seems to 
cause a problem which reminds us of the Hund's 
paradox on the chiral molecules \cite{hund,wightman}. 

The Hamiltonian that this paper deals with reads: 
\begin{equation}
H_{\mathrm{Rabi}}=H_{\mathrm{SS}}
+\hbar\mathrm{g}\left( a+a^{\dagger}\right)\sigma_{x} 
\label{eq:Rabi-Hamiltonian}
\end{equation}
because we assumed the condition 
$\omega:=\omega_{\mathrm{a}}=\omega_{\mathrm{c}}$. 
It is well known \cite{balantekin,SM91,KNT85} 
that we can give a basis 
of a non-compact orthosymplectic superalgebra 
as:
\begin{align*}
& 
K_{+}:=\frac{1}{2}a^{\dagger}a^{\dagger},\,\,\, 
K_{-}:=\frac{1}{2}aa,\,\,\,  
K_{0}:=\frac{1}{4}a^{\dagger}a+\frac{1}{4},\,\,\,  
B:=\frac{1}{4}\sigma_{z}, \\  
& 
W^{\mathrm{R}}_{+}:=\frac{1}{\sqrt{2}}a^{\dagger}\sigma_{-},\,\,\,  
W^{\mathrm{R}}_{-}:=\frac{1}{\sqrt{2}}a\sigma_{+},\,\,\,  
W^{\mathrm{CR}}_{+}:=\frac{1}{\sqrt{2}}a^{\dagger}\sigma_{+},\,\,\,  
W^{\mathrm{CR}}_{-}:=\frac{1}{\sqrt{2}}a\sigma_{-}.       
\end{align*}
Then, our supersymmetric Hamiltonian $H_{\mathrm{SS}}$ is written 
by 
$$
H_{\mathrm{SS}}=2\hbar\omega K_{0}+2\hbar\omega B, 
$$ 
and the operators $W_{\mathrm{R}}$ and $W_{\mathrm{CR}}$, 
respectively called the \textit{rotating term} and 
the \textit{counter-rotating term} in quantum optics, 
are given by 
$$
W_{\mathrm{R}}:= \sqrt{2}(W^{\mathrm{R}}_{+}+W^{\mathrm{R}}_{-})\,\,\,\text{and}
\,\,\, 
W_{\mathrm{CR}}:= \sqrt{2}(W^{\mathrm{CR}}_{+}+W^{\mathrm{CR}}_{-}).
$$ 
They are respectively given by the spin-chiral transformation 
of each other:
$$
W_{\mathrm{CR}}=\sigma_{x}W_{\mathrm{R}}\sigma_{x}.
$$ 
While the rotating term $W_{\mathrm{R}}$ acts 
in the standard state space $\mathcal{F}\otimes\mathbb{C}^{2}$, 
the counter-rotating term $W_{\mathrm{CR}}$ 
becomes a rotating term acting in the chiral state space 
$\mathcal{F}\otimes\sigma_{x}\mathbb{C}^{2}$ 
which is, of course, mathematically 
equal to $\mathcal{F}\otimes\mathbb{C}^{2}$ itself. 
We can rewrite the Rabi Hamiltonian as: 
\begin{equation}
H_{\mathrm{Rabi}}=H_{\mathrm{SS}}+\hbar\mathrm{g}(W_{\mathrm{R}}+W_{\mathrm{CR}})
=H_{\mathrm{SS}}+\hbar\mathrm{g}(W_{\mathrm{R}}+\sigma_{x}W_{\mathrm{R}}\sigma_{x}). 
\label{eq:Rabi-Hamiltonian'}
\end{equation}
The individual contributions from the rotating term $W_{\mathrm{R}}$
and the counter-rotating term $W_{\mathrm{CR}}$ 
(i.e., the chiral rotating term $\sigma_{x}W_{\mathrm{R}}\sigma_{x}$) 
are equitable in the interaction of the Rabi Hamiltonian. 

As shown in the previous section, the coupling-strength 
increase of the Rabi model gives the process from 
the $N=2$\, SUSY system to the system with 
spontaneous SUSY braking caused by the spin-chirality. 
This spin-chirality, in addition, 
shows us another interesting aspect 
in the process. 
The following fact is according to the experimental 
results of circuit QED \cite{HR,RBH,Wallraff08,Gross-Mooij}: 
In the weak and strong coupling regimes of circuit QED, 
the RWA works and thus the so-called 
Jaynes-Cummings (JC) Hamiltonian is 
useful to approximate the Rabi Hamiltonian 
\cite{HR,RBH,Wallraff08} 
in spite of breaking the original equitableness (\ref{eq:Rabi-Hamiltonian'}). 
On the other hand, the effect of the counter-rotating term 
remarkably appears and plays an important role 
when the coupling strength plunges into a region 
beyond that strong coupling regime \cite{Gross-Mooij}, 
while it does not appear so much 
in the strong coupling regime \cite{HR,RBH,Wallraff08}. 
The region beyond the strong coupling regime 
is called the \textit{ultra-strong coupling regime} 
in circuit QED \cite{Gross-Mooij,DGS,CW-ciuti}. 
Namely, the division between the regimes 
of strong and ultra-strong couplings forms 
the division between the validity and the limit of the RWA. 
The present technology of circuit QED has been 
beginning to show us the division. 
Their results say that 
the equitableness (\ref{eq:Rabi-Hamiltonian'}) is broken 
in the weak and strong coupling regimes, 
but the growth of the coupling strength tries 
to recover the equitableness in the ultra-strong coupling regime. 

In this paper we will handle this phenomena from 
the point of view of the CQPT 
\cite{BM07-08} caused by the spin-chirality. 
Then, we follow the classification of the coupling strength regime 
defined by Casanova \textit{et al}. \cite{Solano10}; 
the weak and strong coupling regimes 
are the region between the strengths of 
$\mathrm{g}/\omega=0$ and $\mathrm{g}/\omega=0.1$, 
the ultra-strong coupling regime the region between 
the strengths of $\mathrm{g}/\omega=0.1$ 
and $\mathrm{g}/\omega=1$. 
In addition, the region of the coupling strength 
satisfying the condition $\mathrm{g}/\omega>1$ 
is called the \textit{deep-strong coupling regime}.

The JC Hamiltonian $H_{\mathrm{JC}}$ is obtained by 
applying the RWA to the Rabi Hamiltonian 
and negating the counter-rotating term: 
\begin{equation}
H_{\mathrm{JC}}=H_{\mathrm{SS}}+\hbar\mathrm{g}W_{\mathrm{R}}. 
\label{eq:JC-Hamiltonian}
\end{equation}
In theory we usually assume the conditions:  
\begin{equation}
\mathrm{g}/\omega\ll 1 
\tag{\footnotesize RWA} 
\label{ass:rwa} 
\end{equation}
for the RWA. 
In the case $\omega_{\mathrm{a}}\ne\omega_{\mathrm{c}}$, 
we employ the condition, 
$\mathrm{g}/\omega_{\mathrm{c}}\ll 1 $, 
for the condition (\ref{ass:rwa}). 
In addition  to this, we have to suppose 
another condition 
\begin{equation}
|\omega_{\mathrm{a}}-\omega_{\mathrm{c}}|
\ll\omega_{\mathrm{a}}+\omega_{\mathrm{c}}
\label{eq:well-known-condition-RWA}
\end{equation} 
as well.

For the breaking equitableness and its recovering, 
we focus our attention on the individual roles 
of the rotating term $W_{\mathrm{R}}$ 
and the counter-rotating term $W_{\mathrm{CR}}$. 
Neither the interaction $\hbar\mathrm{g}W_{\mathrm{R}}$ 
nor the interaction $\hbar\mathrm{g}W_{\mathrm{CR}}$ 
can single-handedly make the energy of their own system. 
They need the free energy $H_{\mathrm{SS}}$ 
to pay off. 
Thus, the two interactions scramble for 
the SUSY Hamiltonian $H_{\mathrm{SS}}$ 
in Eq.(\ref{eq:Rabi-Hamiltonian'}) 
to make their individual energy. 
Consequently, as a theoretical attempt, 
it is reasonable to introduce 
another dimensionless parameter $\varepsilon$ 
with $0\le \varepsilon <1$, 
which represents how the interactions, 
$\hbar\mathrm{g}W_{\mathrm{R}}$ and $\hbar\mathrm{g}W_{\mathrm{CR}}$, 
scramble for the SUSY Hamiltonian $H_{\mathrm{SS}}$. 

To introduce this parameter $\varepsilon$ 
in the Rabi Hamiltonian, 
we prepare the two parameterized frequencies 
$\omega_{\mathrm{a}}(\varepsilon)$ and 
$\omega_{\mathrm{c}}(\varepsilon)$, 
and define the parameterized JC Hamiltonian 
$H_{\mathrm{JC}}^{\mathrm{g}}(\varepsilon)$ by 
\begin{equation}
H_{\mathrm{JC}}^{\mathrm{g}}(\varepsilon):=
H_{0}(\varepsilon)
+\hbar\mathrm{g}W_{\mathrm{R}} 
\label{eq:parameterized-JC-Hamiltonian}
\end{equation}
with the parameterized free Hamiltonian: 
\begin{equation}
H_{0}(\varepsilon):=
\frac{\hbar\omega_{\mathrm{a}}(\varepsilon)}{2}\sigma_{z}
+\hbar\omega_{\mathrm{c}}(\varepsilon)
\left(a^{\dagger}a+\frac{1}{2}
\right). 
\label{eq:parameterized-free-Hamiltonian}
\end{equation}
Give our parameterization as: 
$\omega_{\mathrm{a}}(\varepsilon):=
(1+\varepsilon)\omega$ 
and 
$\omega_{\mathrm{c}}(\varepsilon):=
(1-\varepsilon)\omega$.  
Then, the Rabi Hamiltonian is divided into 
the two parts:
\begin{equation}
H_{\mathrm{Rabi}}=
H_{\mathrm{JC}}^{\mathrm{g}}(\varepsilon)
+\varepsilon\sigma_{x}
H_{\mathrm{JC}}^{\mathrm{g}/\varepsilon}(0)\sigma_{x}
\label{eq:chiral-decomposition}
\end{equation} 
for every coupling strength $\mathrm{g}$ 
and the parameter $\varepsilon$ with 
$0\le\mathrm{g}$ and $0<\varepsilon<1$. 
We note that the parameterized JC Hamiltonian 
with the parameter's value, $\varepsilon=0$, 
is the standard JC Hamiltonian: 
$H_{\mathrm{JC}}^{\mathrm{g}}(0)=H_{\mathrm{JC}}$. 
We call the decomposition (\ref{eq:chiral-decomposition}) 
the \textit{chiral decomposition}. 
Here the Hamiltonian $H_{\mathrm{JC}}^{\mathrm{g}/\varepsilon}(0)$ 
is also a parameterized JC Hamiltonian, 
given by replacing the coupling constant 
$\mathrm{g}$ and the parameter $\varepsilon$ 
in Eq.(\ref{eq:parameterized-JC-Hamiltonian}) 
with the scaled coupling constant 
$\mathrm{g}/\varepsilon$ 
and the constant $0$ respectively. 
While the Hamiltonians $H_{\mathrm{Rabi}}$ 
and $H_{\mathrm{JC}}^{\mathrm{g}}(\varepsilon)$ act in 
the state space $\mathcal{F}\otimes\mathbb{C}^{2}$, 
the parameterized JC Hamiltonian 
$H_{\mathrm{JC}}^{\mathrm{g}/\varepsilon}(0)$ 
acts in the chiral state space 
$\mathcal{F}\otimes\sigma_{x}\mathbb{C}^{2}$. 
We call the pair 
$\{ H_{\mathrm{JC}}^{\mathrm{g}}(\varepsilon)\, ,\, 
\varepsilon\sigma_{x}H_{\mathrm{JC}}^{\mathrm{g}/\varepsilon}(0)\sigma_{x}\}$ 
the \textit{chiral pair Hamiltonians} of the Rabi model, 
and moreover, the Hamiltonians $H_{\mathrm{JC}}^{\mathrm{g}}(\varepsilon)$ 
and $\varepsilon\sigma_{x}H_{\mathrm{JC}}^{\mathrm{g}/\varepsilon}(0)\sigma_{x}$ 
the \textit{standard part} and the \textit{chiral part} 
of the chiral pair Hamiltonians, respectively. 
In particular, we call the parameterized JC Hamiltonian 
$H_{\mathrm{JC}}^{\mathrm{g}/\varepsilon}(0)$ 
the \textit{chiral-counter Hamiltonian} for the standard part. 
Since the parameter $\varepsilon$ indicates 
how each of terms $\hbar\mathrm{g}W_{\mathrm{R}}$ 
and $\hbar\mathrm{g}W_{\mathrm{CR}}$ scrambles for 
the SUSY Hamiltonian $H_{\mathrm{SS}}$ in the chiral decomposition, 
the parameter $\varepsilon$ plays a role of a rate 
of the decomposition. 
Thus, we call $\varepsilon$ the \textit{decomposition rate}. 

We will investigate the problem of 
the breaking equitableness and its recovering 
through the chiral decomposition (\ref{eq:chiral-decomposition}). 
To do that we will show some physical properties 
of the parameterized JC model 
in the next section.

\section{GST for Parameterized JC Model}
\label{sec:GST} 

In this section we see how the parameterized JC model 
shows \textit{ground-state transition} (GST). 
Let us set the decomposition rate $\varepsilon$ as 
$0\le\varepsilon\le 1$ throughout this section.

Denote each unit-length eigenstate of 
the parameterized JC Hamiltonian 
$H_{\mathrm{JC}}^{\mathrm{g}}(\varepsilon)$ 
by $\varphi_{\nu}^{\mathrm{g}}(\varepsilon)$, 
and its eigenenergy by $E_{\nu}^{\mathrm{g}}(\varepsilon)$: 
$H_{\mathrm{JC}}^{\mathrm{g}}(\varepsilon)\varphi_{\nu}^{\mathrm{g}}(\varepsilon)
=E_{\nu}^{\mathrm{g}}(\varepsilon)\varphi_{\nu}^{\mathrm{g}}(\varepsilon)$ 
with $\langle \varphi_{\mu}^{\mathrm{g}}(\varepsilon)|
\varphi_{\nu}^{\mathrm{g}}(\varepsilon)\rangle =\delta_{\mu\nu}$ 
for $\nu=0, \pm 1, \pm 2, \cdots$. 
The parameterized JC Hamiltonian has the 
$H_{\mathrm{SS}}$-symmetry: 
$$
[H_{\mathrm{SS}} , H_{\mathrm{JC}}^{\mathrm{g}}(\varepsilon)]=0.
$$  
Consequently, we can completely solve the eigenvalue problem 
for the parameterized JC Hamiltonian. 
This mathematical method was expanded for 
the SUSY-JC model by Alhaidari \cite{alhaidari06}.

Set the integer $\nu$ as $|\nu|:=n+1$ for the photon number $n$. 
Then, we can obtain the concrete expression of each eigenenergy 
and its eigenstate. They are the same expressions 
as in Ref.\cite{BHGS}. 
The eigenenergies are: 
\begin{equation}
\begin{cases}
{\displaystyle 
E_{0}^{\mathrm{g}}(\varepsilon)=
\, -\hbar\Delta_{\varepsilon}/2},  \\ 
{\displaystyle 
E_{+|\nu|}^{\mathrm{g}}(\varepsilon)= 
(1-\varepsilon)\hbar\omega|\nu| 
+\hbar\mathcal{R}_{\nu}^{\mathrm{g}}(\varepsilon)}, \\ 
{\displaystyle 
E_{-|\nu|}^{\mathrm{g}}(\varepsilon)= 
(1-\varepsilon)\hbar\omega|\nu| 
- \hbar\mathcal{R}_{\nu}^{\mathrm{g}}(\varepsilon)},
\end{cases}
\label{eq:eigenenergy-n}
\end{equation} 
where the quantity $\Delta_{\varepsilon}$ is 
the \textit{atom-cavity detuning} given by 
$\Delta_{\varepsilon}:=2\varepsilon\omega$,  
and the quantity $\mathcal{R}_{\nu}^{\mathrm{g}}(\varepsilon)$ 
is given by 
$$
\mathcal{R}_{\nu}^{\mathrm{g}}(\varepsilon)=
\frac{1}{2}\sqrt{\Delta_{\varepsilon}^{2}+4\mathrm{g}^{2}|\nu|\,}  
$$
with the \textit{$n$-photon Rabi frequency} 
$2\mathrm{g}\sqrt{|\nu|}\equiv 2\mathrm{g}\sqrt{n+1}$ 
(see, for example, \S 3.4 of Ref.\cite{HR}). 
In the case $\omega_{\mathrm{a}}\ne\omega_{\mathrm{c}}$, 
we only have to give the atom-cavity detuning 
$\Delta_{\varepsilon}$ by 
$\Delta_{\varepsilon}:=\Delta_{0}
+\varepsilon (\omega_{\mathrm{a}}+\omega_{\mathrm{c}})$ 
with $\Delta_{0}:=\omega_{\mathrm{a}}-\omega_{\mathrm{c}}$. 
All the eigenenergies of the parameterized JC Hamiltonian 
are  $E_{\nu}^{\mathrm{g}}(\varepsilon)$, 
$\nu=0, \pm 1, \cdots$, and there is a relation: 
\begin{equation}
E_{-|\nu|}^{\mathrm{g}}(\varepsilon)
\le E_{+|\nu|}^{\mathrm{g}}(\varepsilon).
\label{eq:-<=+}
\end{equation} 

The concrete expression of each eigenstate 
$\varphi_{\nu}^{\mathrm{g}}(\varepsilon)$ 
corresponding to its eigenenergy 
$E_{\nu}^{\mathrm{g}}(\varepsilon)$ is given as: 
\begin{equation}
\begin{cases}
{\displaystyle 
\varphi_{0}^{\mathrm{g}}(\varepsilon)
= 
|0 , \downarrow\rangle}, \\ 
{\displaystyle 
\varphi_{+|\nu|}^{\mathrm{g}}(\varepsilon) 
=
\cos\theta_{n}^{\mathrm{g}}(\varepsilon)
|n , \uparrow\rangle 
+ \sin\theta_{n}^{\mathrm{g}}(\varepsilon)
|n+1 , \downarrow\rangle}, \\  
{\displaystyle 
\varphi_{-|\nu|}^{\mathrm{g}}(\varepsilon) 
= 
-\, \sin\theta_{n}^{\mathrm{g}}(\varepsilon)
|n , \uparrow\rangle 
+ \cos\theta_{n}^{\mathrm{g}}(\varepsilon)
|n+1 , \downarrow\rangle}, 
\end{cases}
\label{eq:eigenstate-n}
\end{equation} 
where the notation $\theta_{\nu}^{\mathrm{g}}(\varepsilon)$ 
is given by 
$\theta_{\nu}^{\mathrm{g}}(\varepsilon):=
\frac{1}{2}\tan^{-1}
\left( 
2\mathrm{g}\sqrt{|\nu|}/\Delta_{\varepsilon}
\right)$ if $\Delta_{\varepsilon}\ne 0$, 
and $\theta_{\nu}^{\mathrm{g}}(\varepsilon)=\pi/4$ 
if $\Delta_{\varepsilon}=0$. 
We note that the state 
$\varphi_{\pm |\nu|}^{\mathrm{g}}(\varepsilon)$ 
is dressed with $|\nu|-1$ or $|\nu|$ photons at least.

The index $\nu$ runs over all integers 
so that if there is no interaction, 
then the eigenvalue $E_{0}^{0}(\varepsilon)$ 
(i.e., $E_{\nu}^{\mathrm{g}}(\varepsilon)$ 
with $\nu=0$ and $\mathrm{g}=0$) 
becomes the ground-state energy 
(i.e., the inequality 
$E_{0}^{0}(\varepsilon)\le E_{\nu}^{0}(\varepsilon)$ 
holds for any non-zero integer $\nu$). 
As is shown below, however, 
there is such a chance as 
each of eigenenergies $E_{-|\nu|}^{\mathrm{g}}(\varepsilon)$, 
$|\nu|= 1, 2, \cdots$, becomes 
the ground-state energy 
when the coupling strength $\mathrm{g}$ is large. 
This makes many quantum phase transitions 
in Rey's sense \cite{Rey09} 
for the parameterized JC Hamiltonian. 
We call this phenomenum the GST. 

To recognize the GST in brief, 
we consider the following 
mathematical problem. 
First up, we point out that 
the energy $E_{0}^{\mathrm{g}}(\varepsilon)$ 
is a constant function 
of the variable $\mathrm{g}$, 
and the energy 
$E_{\pm|\nu|}^{\mathrm{g}}(\varepsilon)$ 
is almost a first-degree polynomial function 
of the variable $\mathrm{g}$ 
with the asymptotic behaviors: 
$$
\begin{cases}
E_{0}^{\mathrm{g}}(\varepsilon) \sim -\mathrm{g}^{0}
\,\,\, (\text{a negative constant}), \\ 
E_{\pm|\nu|}^{\mathrm{g}}(\varepsilon)\sim \pm\mathrm{g},
\end{cases}
\quad\text{as $\mathrm{g}\to\infty$}.
$$ 
Next, we know that the ground-state energy 
 $E_{\mathrm{JC}}^{\mathrm{g}}(\varepsilon)$ 
of the parameterized JC Hamiltonian 
must be one of eigenenergies, 
$E_{-|\nu|}^{\mathrm{g}}(\varepsilon)$, 
$|\nu|= 0, 1, 2, \cdots$, at least. 
On the other hand, as proved in \S\ref{subsec:proof-JCGSE-estimates}, 
we can show that 
the ground-state energy $E_{\mathrm{JC}}^{\mathrm{g}}(\varepsilon)$ 
satisfies the inequalities: 
\begin{equation}
-\, \frac{\hbar\omega_{\mathrm{a}}(\varepsilon)}{2}
-\hbar\mathrm{g}
-\, \frac{\hbar\mathrm{g}^{2}}{\omega_{\mathrm{c}}(\varepsilon)}  
\le 
E_{\mathrm{JC}}^{\mathrm{g}}(\varepsilon)
\le 
-\, \frac{\hbar\mathrm{g}^{2}}{\omega_{\mathrm{c}}(\varepsilon)}, 
\label{eq:JCGSE-estimates}
\end{equation} 
which implies that 
the ground-state energy $E_{\mathrm{JC}}^{\mathrm{g}}(\varepsilon)$
negatively diverges with the order $2$ as $\mathrm{g}\to\infty$: 
$$
E_{\mathrm{JC}}^{\mathrm{g}}(\varepsilon)
\sim -\mathrm{g}^{2}.
$$ 
We have to explain what has happened 
for the asymptotic behavior of the ground-state 
energy $E_{\mathrm{JC}}^{\mathrm{g}}(\varepsilon)$ 
as $\mathrm{g}\to\infty$ 
and how we can obtain this order $2$. 

\begin{figure}[htbp]
  \begin{center}
  \resizebox{75mm}{!}{\includegraphics{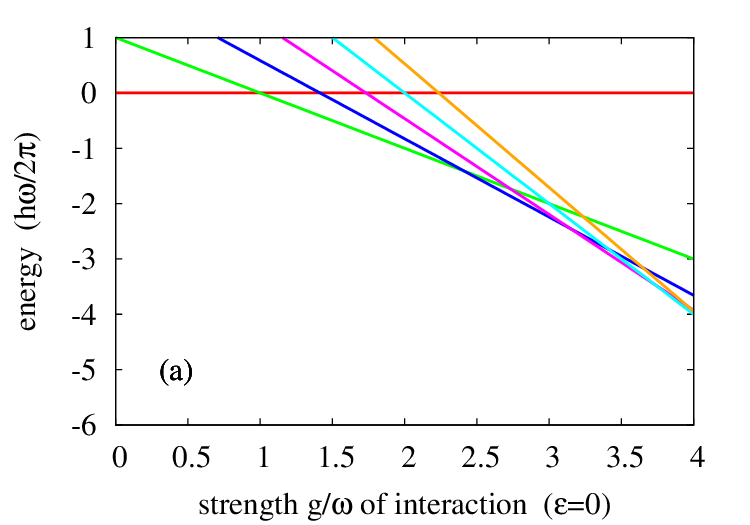}}
\qquad 
  \resizebox{75mm}{!}{\includegraphics{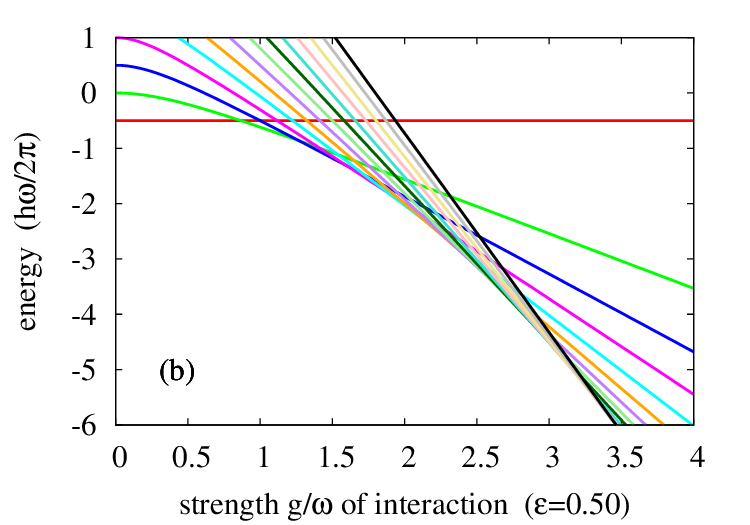}} 
\qquad 
  \resizebox{75mm}{!}{\includegraphics{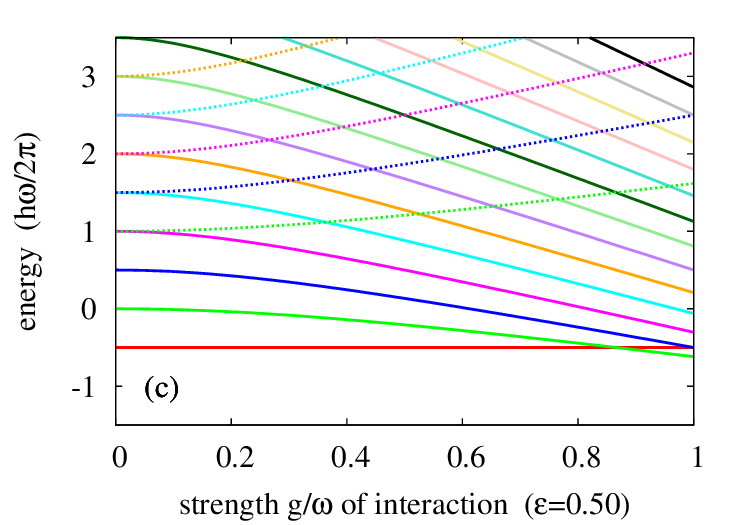}} 
  \end{center}
\vspace*{3mm}
  \caption{\scriptsize 
Energy-level crossings among $E_{\nu}^{\mathrm{g}}(\varepsilon)$, 
$\nu=0, \pm 1, \pm 2, \cdots$, of the parameterized JC Hamiltonian. 
Each color indicates individual index $\nu$ of the energy 
$E_{\nu}^{\mathrm{g}}(\varepsilon)$. 
Here $\omega_{\mathrm{a}}=
\omega_{\mathrm{c}}=\omega$. 
(a) $E_{\nu}^{\mathrm{g}}(\varepsilon)$, 
$\nu=0, -1, -2, \cdots$, of the standard JC model ($\varepsilon=0$); 
(b) $E_{\nu}^{\mathrm{g}}(\varepsilon)$, 
$\nu=0, -1, -2, \cdots$, of the parameterized JC model ($\varepsilon=0.50$); 
(c) $E_{\nu}^{\mathrm{g}}(\varepsilon)$, 
$\nu=0, \pm 1, \pm 2, \cdots$, of the parameterized JC model 
($\varepsilon=0.50$). 
Here, for each $|\nu|$, $E_{\pm|\nu|}^{\mathrm{g}}(\varepsilon)$ are 
drawn with the same color, but 
$E_{+|\nu|}^{\mathrm{g}}(\varepsilon)$ and $E_{-|\nu|}^{\mathrm{g}}(\varepsilon)$ 
are respectively described by the dotted line and the solid one.}
  \label{fig:JC} 
\end{figure}

Actually, as shown in Refs.\cite{hir09-1,hir09-2}, 
when the coupling strength grows larger, 
several energy-level crossings take place 
among energy levels of the JC Hamiltonian. 
For energies $E_{\pm|\nu|}^{\mathrm{g}}(\varepsilon)$, 
$|\nu|= 0, 1, 2, \cdots$, 
see the numerical results in Fig.\ref{fig:JC}.
These energy-level crossings supply us 
with the envelope by the energies, 
which makes the ground-state energy 
with the order $2$. 
This phenomenon is caused by many GSTs:  
We can prove that the eigenstate 
$\varphi_{0}^{\mathrm{g}}(\varepsilon)$
with the eigenenergy $E_{0}^{\mathrm{g}}(\varepsilon)$ 
is the ground state for the coupling strength 
less than a critical coupling strength, 
but the eigenstate $\varphi_{-1}^{\mathrm{g}}(\varepsilon)$  
with the eigenenergy $E_{-1}^{\mathrm{g}}(\varepsilon)$ 
replaces the old ground state $\varphi_{0}^{\mathrm{g}}(\varepsilon)$ 
and becomes the new ground state for the coupling strength 
more than the critical point. 
At the critical point, the parameterized JC Hamiltonian 
has degenerate ground states. 
Each eigenstate $\varphi_{-|\nu|}^{\mathrm{g}}(\varepsilon)$ 
with the energy $E_{-|\nu|}^{\mathrm{g}}(\varepsilon)$, 
$|\nu|= 1, 2, \cdots$, 
also becomes the ground state in turn 
as the coupling strength grows larger and larger, 
though it is primarily an excited state. 
We consequently note that we can also 
find the ground-state entanglement property \cite{PZDS10} 
in this process 
(see \S \ref{sec:CPQT-RabiModel}). 

We make more mathematical statements on the GST here. 
We define the quantity $G_{|\nu|}(\varepsilon)$ by 
\begin{equation}
G_{|\nu|}(\varepsilon):=
(1-\varepsilon)^{2}\omega^{2}
\left[ 
|\nu|+\frac{\Delta_{\varepsilon}}{
\left( 1-\varepsilon\right)\omega}
\right].
\label{eq:quantity-G}
\end{equation} 
Let us now denote by the symbol $\sharp$ 
the equal sign $=$, the inequality sign $>$, 
or the inequality sign $<$. 
Then, in the same way as in Ref.\cite{hir09-1}, 
the direct computation immediately brings us the 
necessary and sufficient condition:
\begin{equation} 
E_{0}^{\mathrm{g}}(\varepsilon)\, \sharp\, 
E_{-|\nu|}^{\mathrm{g}}(\varepsilon)\,\,\, 
\text{if and only if}\,\,\, 
\mathrm{g}^{2}\, \sharp\, 
G_{|\nu|}(\varepsilon).  
\label{eq:criterion1} 
\end{equation} 
In the case $\varepsilon=0$, 
it is easy to compare 
the two energies 
$E_{-|\nu|+1}^{\mathrm{g}}(0)$ and $E_{-|\nu|}^{\mathrm{g}}(0)$: 
\begin{equation}
E_{-|\nu|+1}^{\mathrm{g}}(0)\, \sharp\, 
E_{-|\nu|}^{\mathrm{g}}(0)\quad 
\text{if and only if}\quad 
\mathrm{g}\, \sharp\, 
(\sqrt{|\nu|}+\sqrt{|\nu|-1})\omega
\label{eq:criterion2}
\end{equation}
for $|\nu|= 1, 2, \cdots$.

\section{GST indices and CQPT} 

In this section 
we introduce the \textit{GST indices} (GSTI) 
to see the CQPT for the Rabi model. 

We give the transition probability amplitudes 
$A_{\nu}$ and $B_{\nu}$ by 
$$
\begin{cases}
A_{\nu}:=\langle\varphi_{\nu}^{\mathrm{g}}(\varepsilon)
|\varphi_{\mathrm{Rabi}}\rangle, \\ 
B_{\nu}:=\langle\sigma_{x}\varphi_{\nu}^{\mathrm{g}/\varepsilon}(0)|
\varphi_{\mathrm{Rabi}}\rangle, 
\end{cases}
\nu=0, \pm 1, \pm 2, \cdots, 
$$
for the normalized ground state $\varphi_{\mathrm{Rabi}}$ 
of the Rabi Hamiltonian 
and normalized eigenstates $\varphi_{\nu}^{\mathrm{g}}(\varepsilon)$ 
and $\varphi_{\nu}^{\mathrm{g}/\varepsilon}(0)$ of 
the parameterized JC Hamiltonians. 
As proved in \S\ref{sec:proof-A-B-estimate}, 
the ground-state energy $E_{\mathrm{Rabi}}$ of the Rabi Hamiltonian 
can be expanded as: 
\begin{equation}
E_{\mathrm{Rabi}}=
\sum_{\ell\in\mathbb{Z}}\left[
E_{2\ell}^{\mathrm{g}}(\varepsilon)|A_{2\ell}|^{2}
+\varepsilon E_{2\ell +1}^{\mathrm{g}/\varepsilon}(0)|B_{2\ell +1}|^{2}
\right], 
\label{eq:gse-expansion}
\end{equation}
where $\mathbb{Z}$ denotes the set of all integers. 
We are interested in which term is dominant 
in the expansion (\ref{eq:gse-expansion}). 
If we grasped the transition probabilities 
$|A_{2\ell}|^{2}$ and $|B_{2\ell +1}|^{2}$, 
we could understand how the effect from the chiral space 
contributes to the ground-state energy of the Rabi Hamiltonian. 
But, unfortunately, we have not developed such mathematics 
to grasp them yet. 
Thus, we employ another way. 

We define the lowest-energy sum 
$E_{\mathrm{les}}(\varepsilon)$ 
by the sum of  two ground-state energies 
of chiral pair Hamiltonians of the Rabi model: 
\begin{equation}
E_{\mathrm{les}}(\varepsilon):=
E_{\mathrm{JC}}^{\mathrm{g}}(\varepsilon)
+\varepsilon E_{\mathrm{JC}}^{\mathrm{g}/\varepsilon}(0). 
\label{eq:les}
\end{equation}
Here $E_{\mathrm{JC}}^{\mathrm{g}}(\varepsilon)$ and 
$E_{\mathrm{JC}}^{\mathrm{g}/\varepsilon}(0)$ were respectively 
ground-state energies of the standard part 
$H_{\mathrm{JC}}^{\mathrm{g}}(\varepsilon)$ 
of the chiral pair Hamiltonians and 
its chiral-counter Hamiltonian 
$H_{\mathrm{JC}}^{\mathrm{g}/\varepsilon}(0)$. 
As already explained in \S\ref{sec:GST}, 
the GST takes place for the individual 
parameterized JC Hamiltonians, 
$H_{\mathrm{JC}}^{\mathrm{g}}(\varepsilon)$ 
and $H_{\mathrm{JC}}^{\mathrm{g}_{\varepsilon}}(0)$, 
when the coupling strength is large. 
Thus, we indicate them as 
$E_{\nu_{*}}^{\mathrm{g}}(\varepsilon):=
E_{\mathrm{JC}}^{\mathrm{g}}(\varepsilon)$ 
and 
$E_{\nu_{**}}^{\mathrm{g}/\varepsilon}(0):=
E_{\mathrm{JC}}^{\mathrm{g}/\varepsilon}(0)$
with proper non-positive integers 
$\nu_{*}$ and $\nu_{**}$,  
and we can rewrite the lowest-energy sum as:
\begin{equation}
E_{\mathrm{les}}(\varepsilon)=
E_{\nu_{*}}^{\mathrm{g}}(\varepsilon)
+\varepsilon E_{\nu_{**}}^{\mathrm{g}/\varepsilon}(0). 
\label{eq:E-low}
\end{equation} 
We call the pair $[\, |\nu_{*}|\, ,\, |\nu_{**}|\,](\varepsilon)$ 
of non-negative integers $|\nu_{*}|$ and $|\nu_{**}|$ 
the GSTI for the decomposition rate $\varepsilon$. 
In the case where the ground-state energy 
of the parameterized JC Hamiltonian 
$H_{\mathrm{JC}}^{\mathrm{g}}(\varepsilon)$ has 
some degenerate ground states, 
we employ the non-positive integer $\nu_{*}$ 
so that its absolute value $|\nu_{*}|$ becomes the minimum. 
More precisely, the argument as in \S 2 of Ref.\cite{ah97} 
guarantees that each eigenstate of the parameterized 
JC Hamiltonian is actually unique or finitely degenerate. 
So, we can write 
$\inf\mathrm{Spec}(H_{\mathrm{JC}}^{\mathrm{g}}(\varepsilon))
=E_{\nu_{j}}^{\mathrm{g}}(\varepsilon)$ 
for some $j=1, 2, \cdots, J$. 
Here the notation $\mathrm{Spec}(H)$ stands for 
the set of the energy spectra of a Hamiltonian $H$. 
Thus, we can employ the non-positive integer $\nu_{*}$ 
satisfying $|\nu_{*}|=\min_{j=1, 2, \cdots, J}|\nu_{j}|$ then.  
We adopt the same definition for the 
index $|\nu_{**}|$. 

The GSTI tell us how the GST takes place: 
For non-negative integers $m$ and $n$, 
the change in GSTI from $[m , n](\varepsilon)$ 
to $[m+1 , n](\varepsilon)$ shows that a GST takes place 
for the standard part of the chiral pair Hamiltonians, 
and the change from $[m , n](\varepsilon)$ 
to $[m , n+1](\varepsilon)$ 
means a GST for the chiral part. 
Actually, in the case $\nu_{*}\ne 0$ 
there is the following relation 
among the indices $|\nu_{*}|$ and $|\nu_{**}|$, 
and the decomposition rate $\varepsilon$ 
in the GSTI: 
\begin{equation}
|\nu_{**}|+1\le |\nu_{*}|+\frac{1}{2\varepsilon}, 
\qquad \nu_{*}, \nu_{**}=\, -1, -2, \cdots, 
\label{eq:GSTI-relation}
\end{equation}
which will be proved in \S\ref{subsec:proof-I-III}.   

As proved in \S\ref{subsec:proof-math-statement1}, 
the mathematical statements (\ref{eq:criterion1}) and 
(\ref{eq:criterion2}) lead to 
the following necessary and sufficient condition: 
Define the interval $I_{\varepsilon}$ by 
\begin{equation}
I_{\varepsilon}:=
[0\, ,\, \sqrt{G_{1}(\varepsilon)}]
=[0\, ,\, (1-\varepsilon)\omega
\sqrt{
1+\Delta_{\varepsilon}/(1-\varepsilon)\omega}\,].  
\label{eq:interval}
\end{equation}
Let the decomposition rate $\varepsilon$ be in 
the range between $0$ and $1/2$ 
(i.e, $0\le \varepsilon\le 1/2$), 
and the critical coupling constant $\mathrm{g}[\varepsilon]$ be 
defined by $\mathrm{g}[\varepsilon]:=\varepsilon\omega$. 
Then, for the coupling strength $\mathrm{g}$ running over $I_{\varepsilon}$, 
the critical coupling strength $\mathrm{g}[\varepsilon]$ 
is in the interval $I_{\varepsilon}$, and 
\begin{equation}
\begin{cases}
\text{the GSTI are $[0 , 0](\varepsilon)$ 
if and only if $\mathrm{g}\le\mathrm{g}[\varepsilon]$, 
and moreover,} \\ 
\text{the GSTI are $[0 , n](\varepsilon)$ with some negative 
integer $n$ ($\ne 0$)} \\ 
\text{if and only if $\mathrm{g}>\mathrm{g}[\varepsilon]$.}
\end{cases}
\label{eq:math-statement1}
\end{equation} 
In the case $\omega_{\mathrm{a}}\ne\omega_{\mathrm{c}}$, 
the critical point $\mathrm{g}[\varepsilon]$ 
is actually $\mathrm{g}[\varepsilon]
=\varepsilon\sqrt{\omega_{\mathrm{a}}\omega_{\mathrm{c}}}$. 
The critical point $g[0.5]$ reminds us the critical point 
of the Hepp-Lieb quantum phase transition \cite{HL73,BERB12} 
(see the comment (\ref{eq:numerical-statement}) below).

The chiral decomposition (\ref{eq:chiral-decomposition}) 
says that if the standard part 
$H_{\mathrm{JC}}^{\mathrm{g}}(\varepsilon)$ and 
its chiral part 
$\varepsilon\sigma_{x}
H_{\mathrm{JC}}^{\mathrm{g}_{\varepsilon}}(0)\sigma_{x}$ 
were commutable (mathematically in the sense of 
Definition on p.271 of Ref.\cite{rs1}), 
then the ground-state energy $E_{\mathrm{Rabi}}$ 
of the Rabi Hamiltonian $H_{\mathrm{Rabi}}$ 
would be equal to the lowest-energy sum $E_{\mathrm{les}}(\varepsilon)$. 
But, unfortunately, they are not commutable in fact. 
We therefore define the difference $E_{\mathrm{diff}}(\varepsilon)$ 
between the ground-state energy of the Rabi Hamiltonian 
and the lowest-energy sum by 
$E_{\mathrm{diff}}(\varepsilon)
:=E_{\mathrm{Rabi}}-E_{\mathrm{les}}(\varepsilon)$, 
called \textit{non-commutativity energy}. 
That is, the non-commutativity energy represents 
how the chiral part affects to the standard part 
in the ground-state energy of the Rabi model. 
Conversely, if the effect from the chiral part 
is small, the non-commutativity energy should be small. 
Eq.(\ref{eq:gse-expansion}) leads to the 
expansion of the non-commutativity energy:
$$ 
E_{\mathrm{diff}}(\varepsilon)=
\sum_{\ell\in\mathbb{Z}}\left[
E_{2\ell}^{\mathrm{g}}(\varepsilon)
\left(|A_{2\ell}|^{2}-\delta_{(2\ell)\nu_{*}}\right)
+\varepsilon E_{2\ell +1}^{\mathrm{g}/\varepsilon}(0)
\left(|B_{2\ell +1}|^{2}-\delta_{(2\ell +1)\nu_{**}}\right)
\right]. 
$$
The ground-state energy of the Rabi Hamiltonian is decomposed as:  
\begin{equation}
E_{\mathrm{Rabi}}=
E_{\mathrm{JC}}^{\mathrm{g}}(\varepsilon)
+\varepsilon E_{\mathrm{JC}}^{\mathrm{g}/\varepsilon}(0)
+E_{\mathrm{diff}}(\varepsilon). 
\label{eq:GSE-decomp}
\end{equation}

As a reminder that we follow 
the phenomenology coming from some experimental facts: 
The conditions (\ref{ass:rwa}) implies 
the equitableness breaking, 
and thus, the counter-rotating term is turned on and grows 
as the coupling strength gets larger and larger 
as if to restore the equitableness.  
These experimental facts say in a mathematically naive sense 
that the non-commutativity between the chiral pair 
Hamiltonians should be so small that the standard part 
of the lowest-energy sum plays an important role 
in the weak and strong coupling regime, 
because the chiral-counter Hamiltonian's effect itself is 
too small to show up in the experimental results. 
We cannot, however, ignore it in the ultra-strong coupling regime. 

According to these phenomenological observations, 
since the approximation, 
$E_{\mathrm{Rabi}}\approx E_{\mathrm{JC}}=E_{\mathrm{JC}}^{\mathrm{g}}(0)$, 
experimentally holds for the very small coupling strength 
(i.e., $\mathrm{g}\ll 1$), 
Eq.(\ref{eq:GSE-decomp}) makes us 
expect that:
\begin{equation}
\mathrm{g}\ll 1\, 
\text{implies that}\,\,\, 
\varepsilon\approx 0\,\,\, 
\text{with}\,\,\, 
\varepsilon E_{\mathrm{JC}}^{\mathrm{g}/\varepsilon}(0)
\approx 0\,\,\,\text{and}\,\,\, 
E_{\mathrm{diff}}(\varepsilon)\approx 0. 
\label{eq:our-hope1}
\end{equation}
Moreover, 
\begin{equation}
\begin{cases}
&\text{as the coupling strength $\mathrm{g}$ becomes larger, 
the energy} \\ 
&\text{$\varepsilon E_{\mathrm{JC}}^{\mathrm{g}/\varepsilon}(0)$ must 
appear from the chiral part, and it must} \\ 
&\text{increase gradually.}
\end{cases}
\label{eq:our-hope2}
\end{equation} 
In our arguments below, the statements 
(\ref{eq:our-hope1}) and (\ref{eq:our-hope2}) will be justified. 
Then, the former statement (\ref{eq:our-hope1}) is consistent 
with the RWA. 
The latter statement (\ref{eq:our-hope2}) reveals 
a kind of the CQPT \cite{BM07-08}. 
That is, the CQPT causes the shift of the dominant part of 
the decomposition Eq.(\ref{eq:GSE-decomp}) 
from the standard part of the chiral pair Hamiltonians 
to its chiral part, 
which is represented by GSTI. 
In the process of this shift, it becomes important to grasp 
the behavior of the non-commutativity energy 
$E_{\mathrm{diff}}(\varepsilon)$ as well:  

Our attempt to find the shift 
is not always available for all physical models. 
For example, let $H_{\mathrm{ho}}$ be the $1$-dimensional 
Schr\"{o}dinger operator for the quantum harmonic oscillator: 
$H_{\mathrm{ho}}:=\, -(1/2)d^{2}/dx^{2}
+\mathrm{g}x^{2}$. 
Here we set $\hbar$ as $\hbar=1$ for simplicity. 
Denote the ground-state energy of 
the Schr\"{o}dinger operator $H_{\mathrm{ho}}$ 
by $E_{\mathrm{ho}}$ 
and then it is actually $\sqrt{\mathrm{g}/2}$. 
We set the parameterized Hamiltonian $H_{0}(\varepsilon)$ 
as $H_{0}(\varepsilon):=\, -\varepsilon d^{2}/dx^{2}$. 
We denote its ground-state energy by $E_{0}(\varepsilon)$ 
and then it is actually $0$. 
Here we meant the infimum of energies 
by `ground-state energy' 
though the Hamiltonian $H_{0}(\varepsilon)$ 
does not have a ground state in its state space. 
We have the decomposition 
$H_{\mathrm{ho}}=H_{0}(1/2)+F^{*}H_{0}(\mathrm{g})F$ 
for the Fourier transform $F$. 
The both Hamiltonians $H_{0}(1/2)$ and 
$H_{0}(\mathrm{g})$ are solvable and their 
energies are given 
as $\left[\left. 0,\infty\right)\right.$. 
Define the lowest-energy sum 
$E^{\mathrm{ho}}_{\mathrm{les}}(\varepsilon)$ 
by $E^{\mathrm{ho}}_{\mathrm{les}}(\varepsilon):=
E_{0}(1/2)+E_{0}(\mathrm{g})$. 
Then, we can define the non-commutativity energy 
$E^{\mathrm{ho}}_{\mathrm{diff}}(\varepsilon)$ by 
$E^{\mathrm{ho}}_{\mathrm{diff}}(\varepsilon):=
E_{\mathrm{ho}}-E^{\mathrm{ho}}_{\mathrm{les}}(\varepsilon)$. 
For this model, the lowest-energy sum is actually zero, 
$E^{\mathrm{ho}}_{\mathrm{les}}(\varepsilon)=0$, 
and we have the non-commutativity energy as 
$E^{\mathrm{ho}}_{\mathrm{diff}}(\varepsilon)
=\sqrt{\mathrm{g}/2}$. 
Accordingly, the lowest-energy sum does not make sense 
in the ground-state energy of the Hamiltonian $H_{\mathrm{ho}}$. 
The non-commutativity energy plays an important role 
in the ground-state energy rather than the lowest-energy sum 
for the Hamiltonian $H_{\mathrm{ho}}$. 

Therefore, we have to investigate the behavior of 
the non-commutativity energy for the Rabi model. 

\section{Estimates of Non-Commutativity Energy}

In this section we study the boundedness of 
the non-commutativity energy for the Rabi model. 

\begin{figure}[htbp]
  \begin{center}
  \resizebox{75mm}{!}{\includegraphics{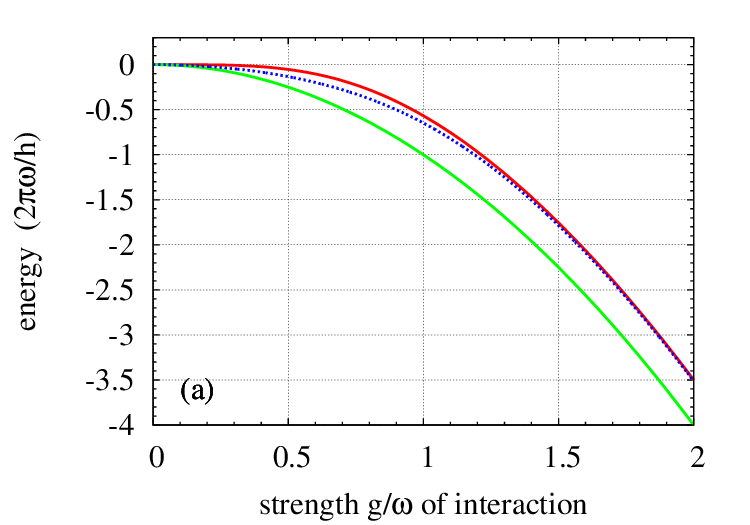}} 
\qquad 
  \resizebox{75mm}{!}{\includegraphics{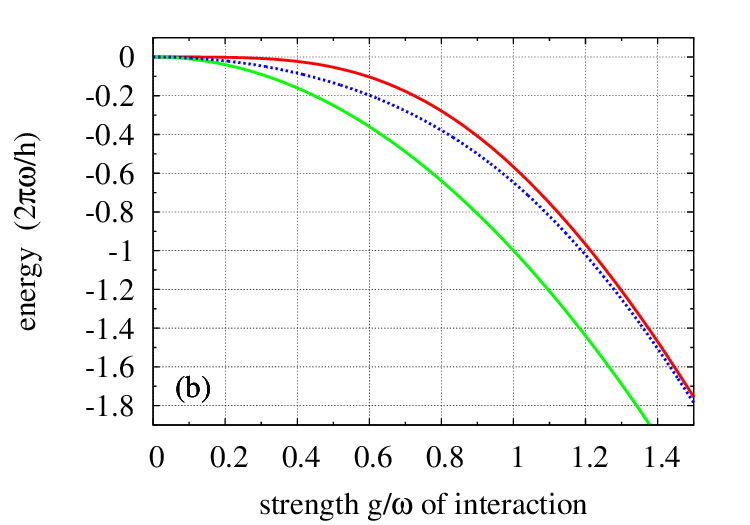}}  \\ 
  \resizebox{75mm}{!}{\includegraphics{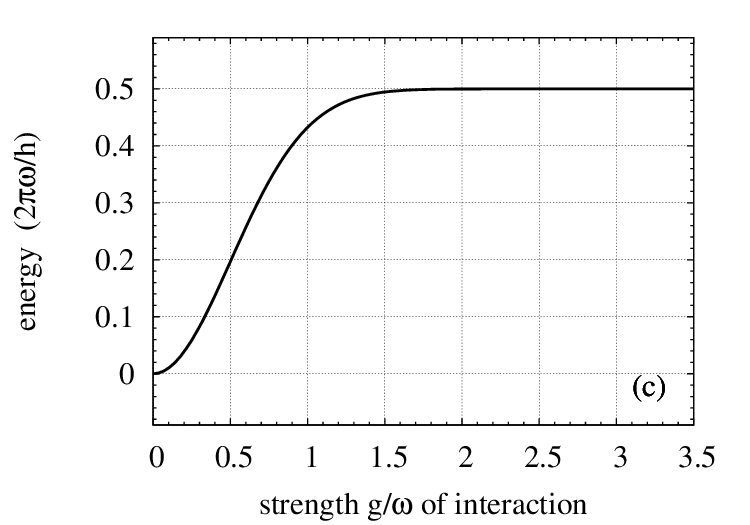}}    
  \end{center}
  \vspace*{3mm}
  \caption{\scriptsize Set $\omega=\omega_{\mathrm{a}}
=\omega_{\mathrm{c}}$. 
$e_{\mathrm{low}}(\mathrm{g})/\hbar\omega$(green solid line), 
$e_{\mathrm{upp}}(\mathrm{g})/\hbar\omega$(red solid line), 
numerically-calculated ground-state energy 
$E_{\mathrm{Rabi}}$(blue dotted line). 
(a) $0\le\mathrm{g}/\omega\le 2$; 
(b) $0\le\mathrm{g}/\omega\le 1.5$; 
(c) $(e_{\mathrm{upp}}(\mathrm{g})
-e_{\mathrm{les}}(\mathrm{g}))/\hbar\omega$(black solid line).   
}
  \label{fig:upper-lower_standard-error-difference-1} 
\end{figure}

For a start, we will give the lower bound 
$E_{\mathrm{lbd}}^{\mathrm{g}}(\varepsilon)$ 
and the upper bound $E_{\mathrm{ubd}}^{\mathrm{g}}(\varepsilon)$ 
of the non-commutativity energy $E_{\mathrm{diff}}(\varepsilon)$ 
to make the estimate: 
\begin{equation}
\max\left\{ 0\, ,\, E_{\mathrm{lbd}}^{\mathrm{g}}(\varepsilon)\right\}
\le E_{\mathrm{diff}}(\varepsilon)\le 
E_{\mathrm{ubd}}^{\mathrm{g}}(\varepsilon),
\label{eq:estimate1}
\end{equation}   
namely, 
\begin{equation}
\max\left\{ 
E_{\mathrm{les}}(\varepsilon)\, ,\, 
E_{\mathrm{les}}(\varepsilon)+E_{\mathrm{lbd}}^{\mathrm{g}}(\varepsilon)
\right\}
\le 
E_{\mathrm{Rabi}} \le 
E_{\mathrm{les}}(\varepsilon)+E_{\mathrm{ubd}}^{\mathrm{g}}(\varepsilon).
\label{eq:estimate1'}
\end{equation} 
So, we determine the lower bound $E_{\mathrm{lbd}}^{\mathrm{g}}(\varepsilon)$ 
and the upper bound $E_{\mathrm{ubd}}^{\mathrm{g}}(\varepsilon)$ now. 
Define two functions $e_{\mathrm{low}}(\mathrm{g})$ 
and $e_{\mathrm{upp}}(\mathrm{g})$ of the variable $\mathrm{g}$ by 
\begin{align*}
& e_{\mathrm{low}}(\mathrm{g}):=\, 
-\, \frac{\hbar\mathrm{g}^{2}}{\omega}, \\ 
& e_{\mathrm{upp}}(\mathrm{g}):=
\frac{\hbar\omega}{2}
\left( 1 - 
e^{-2\mathrm{g}^{2}/\omega^{2}}
\right)  
-\, \frac{\hbar\mathrm{g}^{2}}{\omega}.
\end{align*} 
In the case $\omega_{\mathrm{a}}\ne\omega_{\mathrm{c}}$ 
we set the lower bound $e_{\mathrm{low}}(\mathrm{g})$ 
and the upper bound $e_{\mathrm{upp}}(\mathrm{g})$ as: 
$e_{\mathrm{upp}}(\mathrm{g}):=
(\hbar\omega_{\mathrm{c}}/2)  
-(\hbar\mathrm{g}^{2}/\omega_{\mathrm{c}}) 
-(\hbar\omega_{\mathrm{a}}/2)$  
and 
$e_{\mathrm{upp}}(\mathrm{g}):=
(\hbar\omega_{\mathrm{c}}/2)  
-(\hbar\mathrm{g}^{2}/\omega_{\mathrm{c}}) 
-(\hbar\omega_{\mathrm{a}}/2)
e^{-2\mathrm{g}^{2}/\omega_{\mathrm{c}}^{2}}$. 
Applying variational principle, 
we obtain the estimate: 
\begin{equation}
e_{\mathrm{low}}(\mathrm{g})
\le 
E_{\mathrm{Rabi}}
\le e_{\mathrm{upp}}(\mathrm{g})
\label{eq:estimate2}
\end{equation} 
(See Fig.\ref{fig:upper-lower_standard-error-difference-1}(a) and (b), 
and \S\ref{subsec:proof-estimate2} for its proof). 

Let us make here a small remark. 
As explained in \S\ref{sec:remark-derivation}, 
for the Rabi model we can find some expressions 
similar to those for the instanton gas, 
and then, we can express the ground-state energy $E_{\mathrm{Rabi}}$ 
\`{a} la instanton gas, which also gives 
the estimate (\ref{eq:estimate2}). 

The inequalities (\ref{eq:estimate2}) bring our desired estimate for 
the non-commutativity energy, 
setting the lower bound $E_{\mathrm{lbd}}^{\mathrm{g}}(\varepsilon)$ 
and the upper bound $E_{\mathrm{ubd}}^{\mathrm{g}}(\varepsilon)$ as:
\begin{equation}
\begin{cases}
E_{\mathrm{lbd}}^{\mathrm{g}}(\varepsilon):=
e_{\mathrm{low}}(\mathrm{g})-E_{\mathrm{les}}(\varepsilon), \\ 
E_{\mathrm{ubd}}^{\mathrm{g}}(\varepsilon):=
e_{\mathrm{upp}}(\mathrm{g})-E_{\mathrm{les}}(\varepsilon).  
\end{cases}
\label{eq:lu-bound}
\end{equation}

The inequalities (\ref{eq:estimate1'}) 
lead to the asymptotic behavior:  
\begin{equation}
E_{\mathrm{Rabi}}\approx 
\frac{\hbar\omega}{2}- \hbar\frac{\mathrm{g}^{2}}{\omega}\,\,\, 
\text{as $\mathrm{g}\to\infty$}.
\label{eq:asymp-behavior-*}
\end{equation}
We can estimate the difference between 
the two bounds $e_{\mathrm{low}}(\mathrm{g})$ 
and $e_{\mathrm{upp}}(\mathrm{g})$ as: 
$$
0\le e_{\mathrm{upp}}(\mathrm{g})-e_{\mathrm{low}}(\mathrm{g})
\le
\frac{\hbar\omega}{2}
$$ 
with the limit 
$$
\lim_{\mathrm{g}\to\infty}(e_{\mathrm{upp}}(\mathrm{g})
-e_{\mathrm{low}}(\mathrm{g}))=
\frac{\hbar\omega}{2}
$$ 
(see Fig.\ref{fig:upper-lower_standard-error-difference-1}(c)). 
Namely, the difference between the two bounds, 
$e_{\mathrm{low}}(\mathrm{g})$ and $e_{\mathrm{upp}}(\mathrm{g})$, 
is less than or equal to the zero-point energy 
(i.e., the vacuum fluctuation).

By practically estimating the lower bound 
$E_{\mathrm{lbd}}^{\mathrm{g}}(\varepsilon)$ and 
the upper bound 
$E_{\mathrm{ubd}}^{\mathrm{g}}(\varepsilon)$, 
we can obtain the following estimates of 
the non-commutativity energy as in [I]--[III], 
which will be proved in \S\ref{subsec:proof-I-III}.

[I] For the region $0\le\mathrm{g}/\omega\le\varepsilon<0.1$ 
implying the GSTI $[ 0 , 0](\varepsilon)$, 
we obtain the estimates:  
\begin{equation}
0<\varepsilon\hbar\omega
-\frac{\hbar\mathrm{g}^{2}}{\omega}
\le E_{\mathrm{diff}}(\varepsilon)
\le 
\frac{\hbar\omega}{2}(1-e^{-2\mathrm{g}^{2}/\omega^{2}})
+\varepsilon\hbar\omega
-\frac{\hbar\mathrm{g}^{2}}{\omega}.   
\label{eq:lu-estimate-ws}
\end{equation}
Thus, taking the decomposition rate $\varepsilon$ 
as $\varepsilon=\mathrm{g}/\omega$ leads to
\begin{equation}
E_{\mathrm{diff}}(\mathrm{g}/\omega)
\le 0.1\hbar\omega  
\label{eq:estimate-0.0} 
\end{equation}
at most.

[II] For the GSTI $[0,1](\varepsilon)$ 
we have the estimate as:
\begin{equation}
\max\left\{ 0\, ,\, 
\hbar\mathrm{g}
-\frac{\hbar\mathrm{g}^{2}}{\omega}
\right\}
\le E_{\mathrm{diff}}(\varepsilon)
\le 
\frac{\hbar\omega}{2}(1-e^{-2\mathrm{g}^{2}/\omega^{2}})
+\hbar\mathrm{g}
-\frac{\hbar\mathrm{g}^{2}}{\omega}.   
\label{eq:lu-estimate-u1}
\end{equation}   
This estimate implies at most  
\begin{equation}
E_{\mathrm{diff}}(\varepsilon)
\le 0.53\hbar\omega. 
\label{eq:estimate-0.1}
\end{equation}

[III] For for the GSTI $[\, |\nu_{*}|\, ,\, |\nu_{**}|\,](\varepsilon)$ 
with $|\nu_{*}|\le |\nu_{**}|+1$ for negative indices 
$\nu_{*}, \nu_{**}=\, -1, -2, \cdots$, 
we can show the following estimate:
\begin{align}
& \max\Bigl\{ 0\, ,\, 
\hbar\omega
\Bigl[
-|\nu_{*}|
-\varepsilon(|\nu_{**}|-|\nu_{*}|)
\notag \\ 
&\qquad\qquad\qquad\qquad\qquad 
+\sqrt{
\varepsilon^{2}
+\frac{\mathrm{g}^{2}}{\omega^{2}}
|\nu_{*}|} 
+\frac{\mathrm{g}}{\omega}
\sqrt{|\nu_{**}|}
-\, \frac{\mathrm{g}^{2}}{\omega^{2}}
\Bigr]
\Bigr\}
\notag \\ 
\le&  
E_{\mathrm{diff}}(\varepsilon) 
\notag \\ 
\le& 
\hbar\omega
\Bigl[
\frac{1}{2}-|\nu_{*}|
-\varepsilon(|\nu_{**}|-|\nu_{*}|) 
\notag \\ 
&\qquad\qquad\qquad\qquad 
+\sqrt{
\varepsilon^{2}
+\frac{\mathrm{g}^{2}}{\omega^{2}}
|\nu_{*}|}
+\frac{\mathrm{g}}{\omega}
\sqrt{|\nu_{**}|}
-\, \frac{\mathrm{g}^{2}}{\omega^{2}}
-\, \frac{1}{2}e^{-2\mathrm{g}^{2}/\omega^{2}}
\Bigr].
\label{eq:lu-estimate-u2}
\end{align}   
The upper bound follows from this estimate as:    
\begin{equation}
E_{\mathrm{diff}}(\varepsilon)
\le 0.5\hbar\omega. 
\label{eq:estimate-*.**} 
\end{equation} 
Combining the relation (\ref{eq:GSTI-relation}) 
with the condition for the statement [III], 
we realize the relation:  
\begin{equation}
0< |\nu_{*}|\le |\nu_{**}|+1\le |\nu_{*}|+\frac{1}{2\varepsilon}. 
\label{eq:GSTI-relation-III}
\end{equation}

\begin{figure}[htbp]
  \begin{center}
  \resizebox{75mm}{!}{\includegraphics{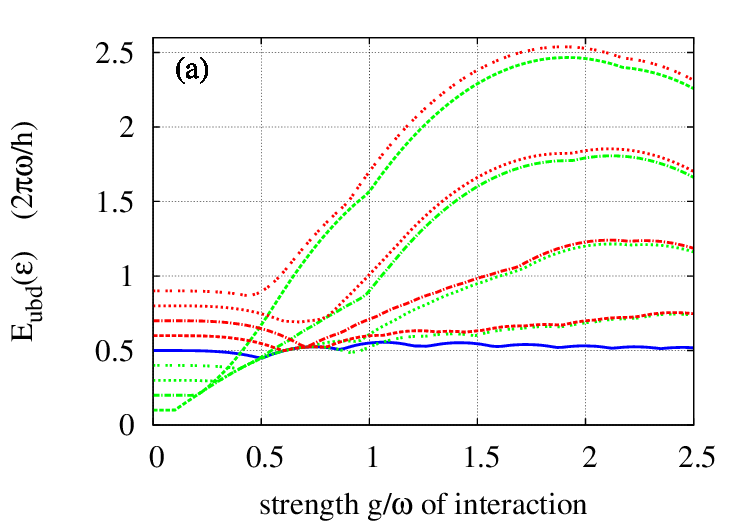}}  
  \resizebox{75mm}{!}{\includegraphics{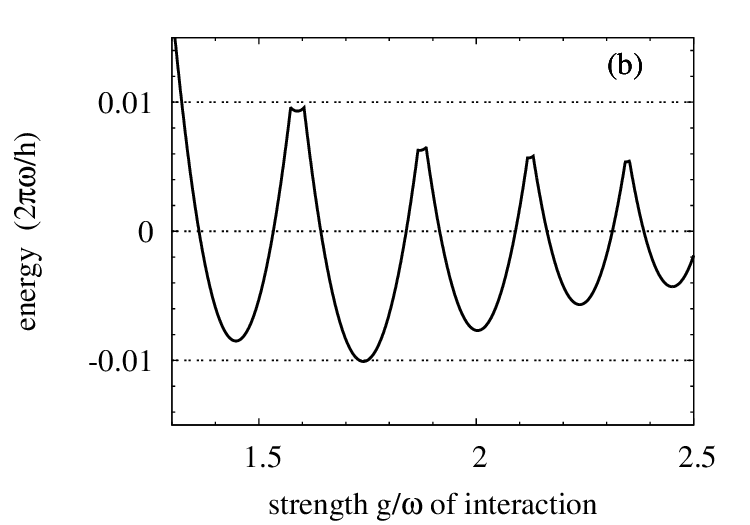}} 
  \end{center}
  \vspace{3mm}
  \caption{\scriptsize 
(a) $E_{\mathrm{ubd}}^{\mathrm{g}}(\varepsilon)/\hbar\omega$. 
From the bottom at $\mathrm{g}/\omega=0$, 
$\varepsilon=0.10$(green dashed line); 
$\varepsilon=0.20$(green dashed-dotted line); 
$\varepsilon=0.30$(green dotted line); 
$\varepsilon=0.40$(green dotted-dotted line); 
$\varepsilon=0.50$(blue solid line); 
$\varepsilon=0.60$(red dashed line); 
$\varepsilon=0.70$(red dashed-dotted line); 
$\varepsilon=0.80$(red dotted line); 
$\varepsilon=0.90$(red dotted-dotted line). 
The envelope of all lines gives a better value 
of $E_{\mathrm{ubd}}^{\mathrm{g}}(\varepsilon)/\hbar\omega$. 
(b)  $(E_{\mathrm{les}}(0.5)
+0.51\hbar\omega-E_{\mathrm{Rabi}})/\hbar\omega$ 
(black solid line), 
where $E_{\mathrm{Rabi}}$ is 
the numerically computed ground-state energy.}
\label{fig:error-estimates-0} 
\end{figure}

The upper bound 
$E_{\mathrm{ubd}}^{\mathrm{g}}(\varepsilon)$ 
is numerically estimated 
as in Fig.\ref{fig:error-estimates-0}(a). 
The numerical results in Fig.\ref{fig:error-estimates-0} 
say that the non-commutativity energy 
$E_{\mathrm{diff}}(\varepsilon)$ is almost 
bounded by the fluctuation of vacuum from above 
for the coupling strength 
less than $\mathrm{g}/\omega=2.5$. 
In particular, the numerical analyses in 
Fig.\ref{fig:error-estimates-0}(b) 
say that we can make the difference between 
the two energies, $E_{\mathrm{Rabi}}$ 
and $E_{\mathrm{les}}(0.5)+E_{\mathrm{diff}}(\varepsilon)$, 
is almost less than $0.01\hbar\omega$ 
if we take the non-commutativity energy 
$E_{\mathrm{diff}}(\varepsilon)$ 
as in Fig.\ref{fig:error-estimates-0}(b) 
for the region $0.5\precsim\mathrm{g}/\omega\precsim 2.5$.
Here we point out that 
\begin{equation}
\begin{cases}
&\text{the value of the decomposition rate, $\varepsilon=0.5$, 
means that} \\ 
&\text{the chiral pair Hamiltonians share the SUSY Hamiltonian} \\ 
&\text{$H_{\mathrm{SS}}$ with each other fairly 
since $\omega_{\mathrm{a}}(0.5)=\omega_{\mathrm{c}}(0.5)
=0.5\omega$.}
\end{cases}
\label{eq:numerical-statement}
\end{equation} 

In addition to an upper bound 
of the non-commutativity energy, 
we have a lower bound from our numerical result: 
Fig.\ref{fig:upper-lower_standard-error-difference-1}(b) 
gives us the inequality $E_{\mathrm{ubd}}^{\mathrm{g}}(\varepsilon)
-E_{\mathrm{diff}}(\varepsilon)=
e_{\mathrm{upp}}(\mathrm{g})-E_{\mathrm{Rabi}}\precsim 0.1\hbar\omega$. 
Since the non-commutativity energy is non-negative 
(e.g., see the last inequality in \S\ref{subsec:proof-I-III}), 
we obtain the lower bound as 
$\max\{ 0 , E_{\mathrm{ubd}}^{\mathrm{g}}(\varepsilon)
-0.1\hbar\omega\} \precsim 
E_{\mathrm{diff}}(\varepsilon)$.

\section{Settlement of Decomposition Rate} 

Introducing the decomposition rate $\varepsilon$ gives the degree of 
freedom of the curvature to the energy-curve of 
the parameterized JC Hamiltonian. 
For example, compare the ground-state energy curves in 
Fig.\ref{fig:JC}.  
In fact, we obtain too many degrees of freedom of the curvature 
to determine the best decomposition rate $\varepsilon$. 
Judging from the condition (\ref{eq:our-hope1}) and 
the boundedness in [I]--[III], 
one of candidates of the best decomposition rate 
may be the minimizer $\varepsilon_{*}$: 
$$
E_{\mathrm{diff}}(\varepsilon_{*})=\inf_{0<\varepsilon<1}E_{\mathrm{diff}}(\varepsilon).
$$ 
Unfortunately there has not yet been an answer 
to this problem on seeking the minimizer $\varepsilon_{*}$.   
It is thus important to find a better decomposition rate $\varepsilon$ 
as far as we can at the present stage 
in order that we study the CQPT in the next section. 
We here propose a temporary criterion: 
When fixing the coupling constant $\mathrm{g}$ arbitrarily, 
we chose a decomposition rate $\varepsilon$ 
so that the non-commutativity energy $E_{\mathrm{diff}}(\varepsilon)$ 
becomes as small as possible.

By our results in mathematics or numeral analysis 
argued in preceding sections, 
we can take a better concrete decomposition rate $\varepsilon$ 
as $\varepsilon=\mathrm{g}/\omega$ 
for the weak and strong coupling regimes 
because any decomposition rate $\varepsilon$ 
satisfies $0\le\mathrm{g}/\omega\le\varepsilon<0.1$ 
in these regimes. 
When the coupling constant $\mathrm{g}$ 
is arbitrarily given in the ultra-strong coupling regime, 
let us chose a better decomposition rate $\varepsilon$ 
among the candidates 
$\mathcal{E}=\left\{ 0.10, 0.20, 
\cdots, 0.90\right\}$. 
This is because we do not have any concrete expression of 
the non-commutativity energy $E_{\mathrm{diff}}(\varepsilon)$, 
and cannot solve the mathematical problem 
on the minimizer $\varepsilon_{*}$ yet. 
We employ the decomposition rate $\varepsilon$ 
by the equation 
\begin{equation}
E_{\mathrm{ubd}}^{\mathrm{g}}(\varepsilon)
=\min_{\varepsilon'\in\mathcal{E}}E_{\mathrm{ubd}}^{\mathrm{g}}(\varepsilon')
\label{eq:temporary-ubd}
\end{equation}
instead.  
We obtain the concrete decomposition rates $\varepsilon$ 
as in Table \ref{fig:epsilon-concrete}. 
Of course, we can find a better candidate $\varepsilon$ 
than ours if we enlarge the set $\mathcal{E}$. 
\begin{center}
\begin{table}[htbp]
\begin{center}
  \begin{tabular}{|c|c|}   \hline 
     {\scriptsize range of coupling strength} 
& {\scriptsize value of decomposition rate}  \\  
     {\scriptsize $\mathrm{g}/\omega$} 
& {\scriptsize $\varepsilon$}  \\ \hline 
    {\scriptsize $0\le\mathrm{g}/\omega\le\varepsilon<0.1$}  
    & {\scriptsize $\mathrm{g}/\omega$}  \\ \hline 
     {\scriptsize $0.1\le\mathrm{g}/\omega<0.2414$} 
    & {\scriptsize $0.10$}  \\ \hline 
     {\scriptsize $0.2414\le\mathrm{g}/\omega<0.4$}
    & {\scriptsize $0.20$}  \\ \hline 
     {\scriptsize $0.4\le\mathrm{g}/\omega<0.5$}
    & {\scriptsize $0.30$}  \\ \hline 
     {\scriptsize $0.5\le\mathrm{g}/\omega<0.9165$}
    & {\scriptsize $0.40$}   \\ \hline 
     {\scriptsize $0.9165\le\mathrm{g}/\omega<0.9659$}
    & {\scriptsize $0.40$}  \\ \hline
     {\scriptsize $0.9659\le\mathrm{g}/\omega<1.193$}
    & {\scriptsize $0.50$}   \\ \hline 
  \end{tabular}
\end{center}
  \caption{\scriptsize The decomposition rate $\varepsilon$ 
defined by Eq.(\ref{eq:temporary-ubd}).}
\label{fig:epsilon-concrete}
\end{table}
\end{center}

\section{CQPT in Rabi Model} 
\label{sec:CPQT-RabiModel}

In this section we investigate how the CQPT 
in the Rabi model takes place 
using the concrete decomposition rate $\varepsilon$ obtained in 
the preceding section.

To begin with, we compute the GSTI for 
each decomposition rate 
$\varepsilon$ corresponding the individual 
coupling strength. 
Then, we can obtain the ground-state energies of 
the chiral pair Hamiltonians 
as in Table \ref{fig:hyou-matome}. 
\begin{center}
\begin{table}[htbp]
\begin{center}
  \begin{tabular}{|c|c|c|c|}   \hline 
     {\scriptsize coupling strength} 
& {\scriptsize GSTI}
& {\scriptsize standard part energy} 
& {\scriptsize chiral part energy}  \\  
     {\scriptsize $\mathrm{g}/\omega$} 
& {\scriptsize $[\, |\nu_{*}|\, ,\, |\nu_{**}|\, ](\varepsilon)$}
& {\scriptsize $E_{\nu_{*}}^{\mathrm{g}}(\varepsilon)$} 
& {\scriptsize $\varepsilon
E_{\nu_{**}}^{\mathrm{g}/\varepsilon}(0)$}  \\ \hline 
    {\scriptsize $0\le\mathrm{g}/\omega\le\varepsilon<0.1$}  
    & {\scriptsize $[0,0](\mathrm{g}/\omega)$}  
    & {\scriptsize $-\hbar\mathrm{g}$}
    & {\scriptsize $0$}  \\ \hline 
     {\scriptsize $0.1\le\mathrm{g}/\omega<0.2414$} 
    & {\scriptsize $[0,1](0.10)$}  
    & {\scriptsize $-0.10\hbar\omega$}
    & {\scriptsize $0.10\hbar\omega-\hbar\mathrm{g}$} \\ \hline 
     {\scriptsize $0.2414\le\mathrm{g}/\omega<0.4$}
    & {\scriptsize $[0,1](0.20)$}  
    & {\scriptsize $-0.20\hbar\omega$} 
    & {\scriptsize $0.20\hbar\omega-\hbar\mathrm{g}$}  \\ \hline 
     {\scriptsize $0.4\le\mathrm{g}/\omega<0.5$}
    & {\scriptsize $[0,1](0.30)$}  
    & {\scriptsize $-0.30\hbar\omega$} 
    & {\scriptsize $0.30\hbar\omega-\hbar\mathrm{g}$} \\ \hline 
     {\scriptsize $0.5\le\mathrm{g}/\omega<0.9165$}
    & {\scriptsize $[0,1](0.40)$}  
    & {\scriptsize $-0.40\hbar\omega$} 
    & {\scriptsize $0.40\hbar\omega-\hbar\mathrm{g}$}  \\ \hline 
     {\scriptsize $0.9165\le\mathrm{g}/\omega<0.9659$}
    & {\scriptsize $[1,1](0.40)$}  
    & {\scriptsize $0.6\hbar\omega
        -\hbar\sqrt{0.16\omega^{2}+\mathrm{g}^{2}}$} 
    & {\scriptsize $0.40\hbar\omega-\hbar\mathrm{g}$}  \\ \hline
     {\scriptsize $0.9659\le\mathrm{g}/\omega<1.193$}
    & {\scriptsize $[1,1](0.50)$}  
    & {\scriptsize $0.5\hbar\omega
        -\hbar\sqrt{0.25\omega^{2}+\mathrm{g}^{2}}$} 
    & {\scriptsize $0.50\hbar\omega-\hbar\mathrm{g}$} \\ \hline 
  \end{tabular}
\end{center}
  \caption{\scriptsize GSTI and ground-state energies of 
Hamiltonians of chiral pair.}
\label{fig:hyou-matome}
\end{table}
\end{center}
The results in Table \ref{fig:hyou-matome} say that 
CQPT takes place when the coupling regime changes 
from the strong coupling regime to the ultra-strong one. 
In addition to this, we can find another 
transition at around $\mathrm{g}/\omega\approx 1.0$, 
namely, when the coupling strength $\mathrm{g}$ 
almost plunges into the deep-strong coupling regime.

\subsection{Weak and Strong Coupling Regimes}
\label{subsec:ws}

In these coupling regimes, 
the index $|\nu_{*}|$ is zero (i.e., $\nu_{*}=0$) 
according to Table \ref{fig:hyou-matome}.  
So, the ground state of the standard part 
$H_{\mathrm{JC}}^{\mathrm{g}}(\varepsilon)$ 
is a separable state $\varphi_{0}^{\mathrm{g}}(\varepsilon)
\equiv|g,0\rangle\equiv|g\rangle\otimes|0\rangle$ 
in the standard state space. 
The eigenenergy of the state 
$\varphi_{0}^{\mathrm{g}}(\varepsilon)$ is 
$E_{\mathrm{JC}}^{\mathrm{g}}(\varepsilon)=\, 
-\varepsilon\hbar\omega
=\, -\hbar\mathrm{g}$ 
since $\varepsilon=\mathrm{g}/\omega$. 
Although the chiral part 
$\varepsilon\sigma_{x}H_{\mathrm{JC}}^{\mathrm{g}/\varepsilon}(0)\sigma_{x}$ 
also has a separable state $\sigma_{x}\varphi_{0}^{\mathrm{g}/\varepsilon}(0)$ 
due to $\nu_{**}=0$, 
the expansion (\ref{eq:gse-expansion}) says that 
the state $\sigma_{x}\varphi_{0}^{\mathrm{g}/\varepsilon}(0)$ 
in the chiral state space makes no contribution 
in the ground-state energy 
of the Rabi Hamiltonian. 
Actually, the ground-state energy of the chiral part 
is zero in our case: 
$\varepsilon E_{\mathrm{JC}}^{\mathrm{g}/\varepsilon}(0)=0$. 
Thus, the growth by the coupling strength $\mathrm{g}$ 
in the ground-state energy 
$E_{\mathrm{Rabi}}$ appears from standard part only, not from the chiral part. 
Accordingly, the following expression works in these two regimes:  
\begin{equation}  
E_{\mathrm{Rabi}}=E_{\mathrm{JC}}^{\mathrm{g}}(\varepsilon)
+E_{\mathrm{diff}}(\varepsilon)
=E_{\mathrm{JC}}^{\mathrm{g}}(\mathrm{g}/\omega)
+E_{\mathrm{diff}}(\mathrm{g}/\omega).
\label{eq:weak-strong-expression}
\end{equation}
Since the decomposition rate $\varepsilon$ 
is now given by $\varepsilon=\mathrm{g}/\omega$, 
the estimate (\ref{eq:lu-estimate-ws}) says 
that the non-commutativity energy approaches 
to zero as the coupling strength decays: 
\begin{equation}
\lim_{\mathrm{g}\to 0}E_{\mathrm{diff}}(\varepsilon)=
\lim_{\mathrm{g}\to 0}E_{\mathrm{diff}}(\mathrm{g}/\omega)=0. 
\label{eq:weak-strong-expression'}
\end{equation} 
Eqs.(\ref{eq:weak-strong-expression}) and 
(\ref{eq:weak-strong-expression'}) give 
a mathematical justification for the RWA 
so that the ground state of 
the Rabi Hamiltonian $H_{\mathrm{Rabi}}$ 
is approximated by that of the parameterized 
JC Hamiltonian $H_{\mathrm{JC}}^{\mathrm{g}}(\varepsilon)$, 
the standard part of the chiral pairs, 
in the weak and strong coupling regimes. 

As for the ground state itself, we can make 
the following argument on the transition probability amplitude $A_{0}$. 
By using a mathematical technique \cite{ah97}, 
we can show the following estimate: 
\begin{equation}
1-\frac{\mathrm{g}^{2}}{\omega^{2}}
\le |A_{0}|^{2}
\le 1, 
\label{eq:A-B-estimate}
\end{equation}
of which proof is in \S\ref{sec:proof-A-B-estimate}. 
This implies the limit, 
$\lim_{\mathrm{g}\to 0}|A_{0}|^{2}=1$. 
Meanwhile, the Rabi Hamiltonian $H_{\mathrm{Rabi}}$ 
converges to the free Hamiltonian 
$(\hbar\omega/2)\sigma_{z}+\hbar\mathrm{g}(a^{\dagger}a+1/2)$ 
as $\mathrm{g}\to 0$ 
in the norm resolvent sense. 
Applying Lemma 4.9 of Ref.\cite{ah97} or 
Theorem VIII.23 of Ref.\cite{rs1} 
to this limit, 
we reach the ground-state limit: 
\begin{equation}
\lim_{\mathrm{g}\to 0}\psi_{\mathrm{Rabi}}
=|g,0\rangle\equiv |g\rangle |0\rangle.
\label{eq:ground-state-limit}
\end{equation} 

Based on the arguments above, 
\textit{the dominant part of the ground state 
of the Rabi Hamiltonian is the separable state 
$|g,0\rangle\equiv |g\rangle |0\rangle$ 
in the weak coupling regime}.   
Since the limit 
$\lim_{\mathrm{g}\to 0}E_{\mathrm{JC}}^{\mathrm{g}}(\mathrm{g}/\omega)
=E_{\mathrm{JC}}$ is obtained, 
our method also reestablishes 
the well-known approximation: 
\begin{equation} 
E_{\mathrm{Rabi}}\approx E_{\mathrm{JC}},\quad 
0\le\mathrm{g}/\omega\ll 1,\qquad 
\label{eq:well-known-approx}
\end{equation} 
which is consistent with the validity of the RWA 
under the condition ({\footnotesize RWA}). 

In the case $\omega_{\mathrm{a}}\ne\omega_{\mathrm{c}}$, 
the expressions of $E_{\mathrm{JC}}^{\mathrm{g}}(\varepsilon)$ 
and $\varepsilon E_{\mathrm{JC}}^{\mathrm{g}/\varepsilon}(0)$ 
tell us that we have to add 
another condition (\ref{eq:well-known-condition-RWA}) 
to our arguments above 
in addition to the condition ({\footnotesize RWA}). 
That is, we also need the smallness, 
$\varepsilon |E_{\mathrm{JC}}^{\mathrm{g}/\varepsilon}(0)|\ll 
|E_{\mathrm{JC}}^{\mathrm{g}}(\varepsilon)|$, 
estimating the chiral part to be so small as  
\begin{equation} 
\varepsilon E_{\mathrm{JC}}^{\mathrm{g}/\varepsilon}(0)
=\, 
-\varepsilon\, \frac{\hbar(\omega_{\mathrm{a}}-\omega_{\mathrm{c}})}{2}
\approx 0
\label{eq:smallness}
\end{equation} 
in comparison with 
$$
E_{\mathrm{JC}}^{\mathrm{g}}(\varepsilon)
=\, -\, \frac{\hbar\Delta_{\varepsilon}}{2}
=\, -\, \frac{\hbar(\omega_{\mathrm{a}}-\omega_{\mathrm{c}})}{2} 
- \varepsilon\, \frac{\hbar(\omega_{\mathrm{a}}+\omega_{\mathrm{c}})}{2}. 
$$
Under the condition (\ref{eq:well-known-condition-RWA}), 
we reach the approximation 
$E_{\mathrm{Rabi}}\approx E_{\mathrm{JC}}=\, 
-\hbar(\omega_{\mathrm{a}}-\omega_{\mathrm{c}})/2$ 
for $0\le\mathrm{g}/\omega\ll 1$. 
Therefore, \textit{in our argument, 
the requirement of the condition 
(\ref{eq:well-known-condition-RWA}) comes from 
the smallness of chiral part}. 
In the standard interpretation 
for the RWA, 
the condition (\ref{eq:well-known-condition-RWA}) 
is usually used for neglecting 
the counter-rotating term $W_{\mathrm{CR}}$ 
after approximating the Heisenberg pictures 
$e^{itH_{\mathrm{Rabi}}}We^{-itH_{\mathrm{Rabi}}}$ of 
the rotating and counter-rotating terms, 
$W=W_{\mathrm{R}}, W_{\mathrm{CR}}$,  
for the Rabi Hamiltonian $H_{\mathrm{Rabi}}$ 
by the Heisenberg pictures 
$e^{itH_{\mathrm{SS}}}We^{-itH_{\mathrm{SS}}}$ 
for the free Hamiltonian $H_{\mathrm{SS}}$.

When the coupling strength grows in 
the weak and strong coupling regimes,
the ground-state energy $E_{\mathrm{JC}}$ of 
the original JC Hamiltonian 
is beginning to produce deviation 
from the ground-state energy $E_{\mathrm{Rabi}}$ 
of the Rabi Hamiltonian. 
But, since the ground-state energy of the Rabi Hamiltonian 
still has the expression (\ref{eq:weak-strong-expression}) 
as far as in the strong coupling regime, 
we can make corrections for the deviation 
by the difference between the ground-state 
energies of the original and parameterized 
JC Hamiltonians, and the non-commutativity energy. 
Namely, the deviation is represented as:
\begin{equation}
E_{\mathrm{Rabi}}-E_{\mathrm{JC}}=
(E_{\mathrm{JC}}^{\mathrm{g}}(\varepsilon)-E_{\mathrm{JC}})
+E_{\mathrm{diff}}(\varepsilon)
\label{eq:deviation}
\end{equation}
in the strong coupling regime 
as well as in the weak one. 
\textit{In this stage, we do not have to consider 
the contribution from the chiral part} 
(i.e., the effect of the counter-rotating terms).

\subsection{Ultra-Strong Coupling Regime}
\label{subsec:us}

Table \ref{fig:hyou-matome} says that 
the index $|\nu_{**}|$ of the GSTI 
becomes one (i.e., $\nu_{**}=\, -1$) 
for the region $0.1<\mathrm{g}/\omega<0.9165$. 
That is, the chiral-counter Hamiltonian 
$\varepsilon\sigma_{x}H_{\mathrm{JC}}^{\mathrm{g}/\varepsilon}(0)\sigma_{x}$ 
has a GST, and then, 
its ground state is $\varphi_{-1}^{\mathrm{g}/\varepsilon}(0)
\equiv 
-\, \sin\theta_{0}^{\mathrm{g}/\varepsilon}(0)
|e,0\rangle 
+ \cos\theta_{0}^{\mathrm{g}/\varepsilon}(0)
|g,1\rangle$ 
in the chiral state space, 
which is an entangled state 
and dressed with one photon.  
Therefore, the chiral part energy 
$\varepsilon E_{\mathrm{JC}}^{\mathrm{g}/\varepsilon}(0)$ 
is turned on and works with the expression 
$\varepsilon E_{\mathrm{JC}}^{\mathrm{g}/\varepsilon}(0)
=\varepsilon\hbar\omega
-\hbar\mathrm{g}$. 
This shows how the effect of the chiral part appears. 
On the other hand, 
the ground-state energy $E_{\mathrm{JC}}^{\mathrm{g}}(\varepsilon)$ 
is given by 
$E_{\mathrm{JC}}^{\mathrm{g}}(\varepsilon)=\, 
-\varepsilon\hbar\omega$ 
as before since the index $|\nu_{*}|$ is still zero. 
Thus, the ground-state energy $E_{\mathrm{Rabi}}$ of 
the Rabi Hamiltonian has the expression: 
\begin{equation}
E_{\mathrm{Rabi}}
=E_{\mathrm{JC}}^{\mathrm{g}}(\varepsilon)
+\varepsilon E_{\mathrm{JC}}^{\mathrm{g}/\varepsilon}(0)
+E_{\mathrm{diff}}(\varepsilon)
=\, 
-\hbar\mathrm{g}
+E_{\mathrm{diff}}(\varepsilon)
\label{eq:growth}
\end{equation}
with 
$$
\begin{cases}
E_{\mathrm{JC}}^{\mathrm{g}}(\varepsilon)
=\, -\varepsilon\hbar\omega, \\ 
\varepsilon E_{\mathrm{JC}}^{\mathrm{g}/\varepsilon}(0)
=\varepsilon\hbar\omega-\hbar\mathrm{g}
=\, -E_{\mathrm{JC}}^{\mathrm{g}}(\varepsilon)-\hbar\mathrm{g}, \\ 
0\le E_{\mathrm{diff}}(\varepsilon)\le 
\text{the upper bound in the estimate 
(\ref{eq:lu-estimate-u1})
}, 
\end{cases}
$$
for $0.1\le \mathrm{g}/\omega \precsim 0.9$ 
(See Fig.\ref{fig:error-estimates-0}). 
The energy $\varepsilon E_{\mathrm{JC}}^{\mathrm{g}/\varepsilon}(0)$ 
from the \textit{chiral part} and the non-commutativity energy $E_{\mathrm{diff}}(\varepsilon)$ 
between the standard part and the \textit{chiral part} 
arise and play an essential role. 
The energy $\varepsilon E_{\mathrm{JC}}^{\mathrm{g}/\varepsilon}(0)$ from the chiral part 
kills the energy $E_{\mathrm{JC}}^{\mathrm{g}}(\varepsilon)$ 
from the standard part, 
and it, together with the non-commutativity energy $E_{\mathrm{diff}}(\varepsilon)$, 
makes the effect of the coupling strength $\mathrm{g}$ 
in the ground-state energy $E_{\mathrm{Rabi}}$ 
as follows: $\max\{-\hbar\mathrm{g},-\hbar\mathrm{g}^{2}/\omega\}
\le E_{\mathrm{Rabi}}\le\hbar(1-e^{-2\mathrm{g}^{2}/\omega^{2}})/2
-\hbar\mathrm{g}^{2}/\omega$. 
Here, the energy $E_{\mathrm{JC}}^{\mathrm{g}}(\varepsilon)$ 
from the standard part makes no contribution in the ground-state energy 
of the Rabi model. 
Therefore, we can say that the effect from the chiral part begins to 
appear and takes the initiative in the ground-state energy of the Rabi Hamiltonian 
in the region of $0.1\precsim\mathrm{g}/\omega\precsim 0.9$.

\section{Conclusion} 

We conclude this paper by summarizing how the CQPT 
is caused in the Rabi model by the GSTs 
occurring in the standard part and the chiral part 
of the chiral pair Hamiltonians. 

We have showed that the growth of the coupling strength 
in the Rabi model plays a role of taking the $N=2$\, SUSY 
to the spontaneous SUSY breaking. 
This spontaneous symmetry breaking is caused by 
the spin-chirality. 
In the process of the growth of the coupling strength, 
the spin-chirality makes the CQPT: 
While the contribution 
from the chiral part is 
so small that we can ignore it 
in the weak coupling regime as in Eq.(\ref{eq:well-known-approx}), 
the deviation appears like Eq.(\ref{eq:deviation}) 
when the coupling strength grows in the strong coupling regime. 
The contribution from the chiral part is 
turned on and grows as in Eq.(\ref{eq:growth}) 
in the ultra-strong coupling regime. 
When the coupling strength plunges into 
the ultra-strong coupling regime 
from the strong coupling regime, 
the GST takes place in the chiral part of the Rabi Hamiltonian. 
In association with this GST, 
as explained in \S\ref{sec:CPQT-RabiModel}, 
the ground state of the chiral-counter Hamiltonian changes 
from the separate state $\varphi_{0}^{\mathrm{g}/\varepsilon}(0)$ 
to the entangled state $\varphi_{-1}^{\mathrm{g}/\varepsilon}(0)$. 
The growth in the coupling strength of 
the ground state energy of the Rabi Hamiltonian 
is completely governed by the contribution from 
the chiral part until $\mathrm{g}/\omega\approx 0.9$.  
This is the explanation of the transition from 
the strong coupling regime to the ultra-strong one 
by the CQPT. 
We note that we can also find the almost same 
CQPT properties 
even in the case $\omega_{\mathrm{a}}\ne\omega_{\mathrm{c}}$ 
with the condition (\ref{eq:well-known-condition-RWA}).

\hfill\break 
{\large {\bf Acknowledgment}} 
 
The author acknowledges the support from JSPS, 
Grant-in-Aid for Scientific Research (C) 23540204. 
He expresses special thanks to 
Pierre-Marie Billangeon and Yasunobu Nakamura 
for useful discussions, which aroused the author's interest 
in the problems in this paper. 
He is also grateful to Hans Mooij, Kae Nemoto, 
Kouichi Semba, Enrique Solano, and Tsuyoshi Yamamoto 
for useful discussions on theoretical 
and experimental aspects of circuit QED.

\appendix 

\section{A Mathematical Justification of 
Asymptotic Behavior of Rabi Hamiltonian}
\label{sec:justification}

In this appendix, we mathematically justify 
the asymptotic behavior (\ref{eq:asymp-behavior}). 

For every $z\in\mathbb{C}\setminus\mathbb{R}$, 
we can obtain the difference between the resolvent 
$\left( 
U_{\mathrm{g}}^{*}
\left( H_{\mathrm{Rabi}}+\hbar\mathrm{g}^{2}/\omega\right) 
U_{\mathrm{g}}-z\right)^{-1}$ and the resolvent 
$(\widetilde{H}_{0}-z)^{-1}$ 
by using the second resolvent equation along with 
Eq.(\ref{eq:new-*1}): 
\begin{align*}
& \frac{1}{\widetilde{H}_{0}-z}\, 
-\, \frac{1}{
U_{\mathrm{g}}^{*}
\left( H_{\mathrm{Rabi}}+\hbar\mathrm{g}^{2}/\omega\right) 
U_{\mathrm{g}}-z} \\ 
=&\, 
-\, \frac{1}{\widetilde{H}_{0}-z}
\left(\frac{\hbar\omega}{2}\widetilde{V}_{\mathrm{g}}\right)
\frac{1}{
U_{\mathrm{g}}^{*}
\left( H_{\mathrm{Rabi}}+\hbar\mathrm{g}^{2}/\omega\right) 
U_{\mathrm{g}}-z}. 
\end{align*}
Using this equation, Eq.(\ref{eq:weak-decay}), 
and the fact that the resolvent 
$(\widetilde{H}_{0}-z)^{-1}$ is a compact operator, 
Theorem VI.II of Ref.\cite{rs1} yields the strong 
resolvent convergence: 
\begin{equation}
\lim_{\mathrm{g}\to\infty}
\frac{1}{
U_{\mathrm{g}}^{*}
\left( H_{\mathrm{Rabi}}+\hbar\mathrm{g}^{2}/\omega\right) 
U_{\mathrm{g}}-z}\psi 
= \frac{1}{\widetilde{H}_{0}-z}\psi
\label{eq:new-*2}
\end{equation}
for every wave function $\psi$. 

For any Borel set $\Omega$ of the $1$-dimensional 
Euclidean space $\mathbb{R}$, 
we denote by $P_{\mathrm{Rabi}}(\Omega)$ 
and $P_{0}(\Omega)$ the projection-valued measures 
so that 
\begin{align*}
U_{\mathrm{g}}^{*}
\left( H_{\mathrm{Rabi}}+\hbar\mathrm{g}^{2}/\omega\right) 
U_{\mathrm{g}} 
= \int_{-\infty}^{\infty}\lambda P_{\mathrm{Rabi}}(d\lambda)\quad
\text{and}\quad  
\widetilde{H}_{0} 
= \int_{-\infty}^{\infty}\lambda P_{0}(d\lambda).  
\end{align*} 
We recall the following properties.  
Let us denote $P_{\mathrm{Rabi}}(d\lambda)$ or 
$P_{0}(d\lambda)$ by $P_{\sharp}(d\lambda)$. 
Since both Hamiltonians 
$U_{\mathrm{g}}^{*}
\left( H_{\mathrm{Rabi}}+\hbar\mathrm{g}^{2}/\omega\right) 
U_{\mathrm{g}}$ 
and $\widetilde{H}_{0}$ have only isolated discrete eigenvalues, 
if the finite interval $(\alpha , \beta)$ contains no eigenvalue, 
then $P_{\sharp}((\alpha , \beta))=0$. 
Conversely, if the interval $(\alpha , \beta)$ contains 
some eigenvalues, then the number of the eigenvalues is finite, 
and then, $P_{\sharp}((\alpha , \beta))
=P_{\sharp}(\{ E_{1}, \cdots, E_{\ell}\})$, 
where $E_{1}, \cdots, E_{\ell}$ are the finite eigenvalues.

Let $\left\{\alpha_{m}\right\}_{m=1}^{\infty}$ be 
a sequence satisfying 
$$
\hbar\omega\left(m-\,\frac{1}{2}\right)<\alpha_{m}
<\hbar\omega\left(m+\frac{1}{2}\right).
$$ 
Because of the estimate (\ref{eq:estimate2}), the set 
$[0 , \alpha_{1}]\cup 
\bigcup_{m=1}^{\infty}\left.\left(\alpha_{m} , 
\alpha_{m+1}\right.\right]$ 
covers the set of all the energy levels of the renormalized Rabi Hamiltonian: 
\begin{equation}
\mathrm{Spec}\, (H_{\mathrm{Rabi}}+\hbar\mathrm{g}^{2}/\omega) 
\subset 
[0 , \alpha_{1}]\cup 
\bigcup_{m=1}^{\infty}\left.\left(\alpha_{m} , 
\alpha_{m+1}\right.\right]. 
\label{eq:cover}
\end{equation} 
Here $\mathrm{Spec}\, (H)$ is the set of all energy 
of a Hamiltonian $H$. 
 
For each natural number $m$, 
we take a positive number $\varepsilon_{m}$ 
so that 
$$\hbar\omega\left( m-\,\frac{1}{2}\right)
<\alpha_{m}-\varepsilon_{m}
\quad \text{and}\quad \alpha_{m+1}+\varepsilon_{m}
<\hbar\omega\left(m+1+\frac{1}{2}\right).
$$  
Applying Theorem VIII.24(b) of Ref.\cite{rs1} 
to Eq.(\ref{eq:new-*2}), we have the limit 
\begin{align}
\lim_{\mathrm{g}\to\infty}
P_{\mathrm{Rabi}}((\alpha_{m}-\varepsilon_{m} , 
\alpha_{m+1}+\varepsilon_{m}))\psi
&=P_{0}((\alpha_{m}-\varepsilon_{m} , 
\alpha_{m+1}+\varepsilon_{m}))\psi 
\notag \\ 
&=P_{0}\left(\left\{\hbar\omega
\left(m+\frac{1}{2}\right)\right\}\right)\psi, 
\label{eq:new-*3}
\end{align}
which means that the state $P_{\mathrm{Rabi}}((\alpha_{m} , \alpha_{m+1}))\psi$ 
converges to an eigenstate of the asymptotically free Hamiltonian 
$\widetilde{H}_{0}$. 
Meanwhile, by the contraposition of Theorem VIII.24(a) 
of Ref.\cite{rs1}, we have 
\begin{equation}
(\alpha_{m}-\varepsilon_{m} , 
\alpha_{m+1}+\varepsilon_{m})\cap 
\mathrm{Spec}\, \left( 
U_{\mathrm{g}}^{*}
\left( H_{\mathrm{Rabi}}+\hbar\mathrm{g}^{2}/\omega\right) 
U_{\mathrm{g}} 
\right)
\ne \emptyset\label{eq:new-*4}
\end{equation}
for sufficiently large $\mathrm{g}$ 
because we have  $(\alpha_{m}-\varepsilon_{m} , 
\alpha_{m+1}+\varepsilon_{m})\cap 
\mathrm{Spec}\, (\widetilde{H}_{0})
=\left\{ \hbar\omega\left( m+\frac{1}{2}\right)\right\}$. 

For each eigenstate $\psi^{\mathrm{Rabi}}$ of 
the Hamiltonian $U_{\mathrm{g}}^{*}(H_{\mathrm{Rabi}}
+\hbar\mathrm{g}^{2}/\omega)U_{\mathrm{g}}$, 
there is a natural number $n$ so that 
its eigenenergy is in the interval 
$(\alpha_{n}-\varepsilon_{n} , 
\alpha_{n+1}+\varepsilon_{n})$, 
which is ensured by Eq.(\ref{eq:cover}). 
So, we denote the eigenstate $\psi^{\mathrm{Rabi}}$ 
by $\psi^{\mathrm{Rabi}}_{n}$, i.e., 
$\psi^{\mathrm{Rabi}}_{n}:=\psi^{\mathrm{Rabi}}$. 
Thus, Eqs.(\ref{eq:new-*3}) and (\ref{eq:new-*4}) 
say that 
\begin{align}
\lim_{\mathrm{g}\to\infty}\psi^{\mathrm{Rabi}}
&=\lim_{\mathrm{g}\to\infty}\psi^{\mathrm{Rabi}}_{n} 
\notag \\ 
&=
\lim_{\mathrm{g}\to\infty}
P_{\mathrm{Rabi}}((\alpha_{n}-\varepsilon_{n} , 
\alpha_{n+1}+\varepsilon_{n}))\psi^{\mathrm{Rabi}}_{n} 
\notag \\ 
&=P_{0}\left(\left\{\hbar\omega
\left(n+\frac{1}{2}\right)\right\}\right)
\psi^{\mathrm{Rabi}}_{n}
\ne 0.
\label{eq:new-*5}
\end{align}
Here $P_{0}\left(\left\{\hbar\omega
\left( n+\frac{1}{2}\right)\right\}\right)
\psi^{\mathrm{Rabi}}_{n}$ is an eigenstate of 
the asymptotically free Hamiltonian 
$\widetilde{H}_{0}$. 
Consequently, we can say that 
for each eigenstate $\psi^{\mathrm{Rabi}}$ of 
the Hamiltonian $U_{\mathrm{g}}^{*}(H_{\mathrm{Rabi}}
+\hbar\mathrm{g}^{2}/\omega)U_{\mathrm{g}}$, 
there is an eigenstate $\psi_{n}$ 
of the asymptotically free Hamiltonian 
$\widetilde{H}_{0}$ so that   
\begin{equation}
\lim_{\mathrm{g}\to\infty}\psi^{\mathrm{Rabi}}
=\psi_{n}.  
\label{eq:new-*6}
\end{equation}

Conversely, we can prove that 
for each eigenstate $\psi_{n}$ 
of the asymptotically free Hamiltonian 
$\widetilde{H}_{0}$, 
there is an eigenstate $\psi^{\mathrm{Rabi}}$ of 
the Hamiltonian $U_{\mathrm{g}}^{*}(H_{\mathrm{Rabi}}
+\hbar\mathrm{g}^{2}/\omega)U_{\mathrm{g}}$ so that 
Eq.(\ref{eq:new-*6}) holds 
in the following: 
Eqs.(\ref{eq:cover}) and (\ref{eq:new-*4}) ensure 
that for each $n$ there is an eigenstate 
$\psi^{\mathrm{Rabi}}$ of the Hamiltonian 
$U_{\mathrm{g}}^{*}(H_{\mathrm{Rabi}}
+\hbar\mathrm{g}^{2}/\omega)U_{\mathrm{g}}$ so that its energy 
belongs to the interval $(\alpha_{n}-\varepsilon_{n}  , 
\alpha_{n+1}+\varepsilon_{n})$ 
for sufficiently large coupling strength. 
Thus, we obtain Eq.(\ref{eq:new-*5}) again, 
which implies Eq.(\ref{eq:new-*6}). 

Therefore, we obtain the correspondence (\ref{eq:asymp-behavior}). 
Note Eq.(\ref{eq:parity}) now. 
Then, we realize that we can chose $\psi_{n}$ 
as either of one of eigenvectors,  
$|n,\uparrow\rangle$ and $|n,\downarrow\rangle$, 
of the asymptotically free Hamiltonian $\widetilde{H}_{0}$. 
In addition, we can chose any eigenstate of the Hamiltonian 
$U_{\mathrm{g}}^{*}(H_{\mathrm{Rabi}}
+\hbar\mathrm{g}^{2}/\omega)U_{\mathrm{g}}$ 
which converges to that $\psi_{n}$. 
Moreover, applying Theorem VIII.24(a) of Ref.\cite{rs1}, 
we can derive the relation, 
$$
\lim_{\mathrm{g}\to\infty}
\left( 
E^{\mathrm{Rabi}}+\frac{\hbar\mathrm{g}^{2}}{\omega}
\right)
=\hbar\omega\left( n+\frac{1}{2}\right),
$$
between all the energies $E^{\mathrm{Rabi}}$ 
and all the energies $\hbar\omega\left( 
n+\frac{1}{2}\right)$ 
from Eqs.(\ref{eq:cover}) and (\ref{eq:new-*4}), 
which implies the relation (\ref{eq:asymp-behavior'}).

\section{Remarks on Relation with Instanton Gas}
\label{sec:remark-derivation}

Since the Rabi  model is the one-mode photon version 
of the spin-boson model, 
we can apply several results on the spin-boson 
Hamiltonian to the Rabi Hamiltonian. 
In Ref.\cite{hir99} we gave a strict expression of 
the ground-state energy of the spin-boson model 
using the parity conservation 
between the Rabi Hamiltonian $H_{\mathrm{Rabi}}$ and 
the parity operator $\sigma_{z}(-1)^{a^{\dagger}a}$: 
\begin{equation}
[H_{\mathrm{Rabi}},\sigma_{z}(-1)^{a^{\dagger}a}]=0.
\label{eq:parity-symmetry}
\end{equation} 
The method for the expression in Ref.\cite{hir99} 
reminds us of the computation for seeking 
the transition amplitude of the so-called 
instanton gas by the Euclidean path integral \cite{as10}. 
In this appendix we handle general frequencies, 
$\omega_{\mathrm{a}}$ and $\omega_{\mathrm{c}}$, 
that is, we accept the condition, 
$\omega_{\mathrm{a}}\ne\omega_{\mathrm{c}}$.  

We define functions 
$I_{\mathrm{even}}(\beta)$ and $I_{\mathrm{odd}}(\beta)$ 
of a variable $\beta\ge 0$ by 
\begin{align*}
I_{\mathrm{even}}(\beta):=&
1+\sum_{\ell=1}^{\infty}
\left(\frac{\omega_{\mathrm{a}}}{2}\right)^{2\ell}
\int_{0}^{\beta}\!\!\!\!\!\!{d\beta_{1}}
\int_{0}^{\beta_{1}}\!\!\!\!\!\!{d\beta_{2}}
\cdots\int_{0}^{\beta_{2\ell-1}}\!\!\!\!\!\!{d\beta_{2\ell}} \\ 
&\times 
e^{-2(\mathrm{g}^{2}/\omega_{\mathrm{c}}^{2})
(2G_{\beta_{1},\cdots,\beta_{2\ell}}+2\ell)}, \\   
I_{\mathrm{odd}}(\beta):=&
\beta\frac{\omega_{\mathrm{a}}}{2}
e^{-2(\mathrm{g}^{2}/\omega_{\mathrm{c}}^{2})} \\ 
& 
+\sum_{\ell=1}^{\infty}
\left(\frac{\omega_{\mathrm{a}}}{2}\right)^{2\ell+1}
\int_{0}^{\beta}\!\!\!\!\!\!{d\beta_{1}}
\int_{0}^{\beta_{1}}\!\!\!\!\!\!{d\beta_{2}}
\cdots\int_{0}^{\beta_{2\ell}}\!\!\!\!\!\!{d\beta_{2\ell+1}} \\ 
&\times 
e^{-2(\mathrm{g}^{2}/\omega_{\mathrm{c}}^{2})
(2G_{\beta_{1},\cdots,\beta_{2\ell}}+
2F_{\beta_{1},\cdots,\beta_{2\ell+1}}+\left(2\ell +1\right))} 
\end{align*}
for the sequences 
$\left\{G_{\beta_{1},\cdots,\beta_{2\ell}}\right\}_{\ell=1}^{\infty}$ 
and 
$\left\{F_{\beta_{1},\cdots,\beta_{2\ell+1}}\right\}_{\ell=0}^{\infty}$ 
given by 
\begin{align*}
G_{\beta_{1},\cdots,\beta_{2\ell}}
=& 
\, -\, \sum_{p=1}^{\ell}
e^{-(\beta_{2p-1}-\beta_{2p})\omega_{\mathrm{c}}} \\ 
& + 
\sum_{p, q=1; p<q}^{\ell}
\left(
e^{-\beta_{2p-1}\omega_{\mathrm{c}}} 
- 
e^{-\beta_{2p}\omega_{\mathrm{c}}}
\right) \\ 
&\qquad\qquad\quad
\times\left(
e^{\beta_{2q-1}\omega_{\mathrm{c}}} 
- 
e^{\beta_{2q}\omega_{\mathrm{c}}}
\right) \le 0, 
\end{align*}
and 
\begin{align*} 
F_{\beta_{1},\cdots,\beta_{2\ell+1}}
=& 
e^{\beta_{2\ell+1}\omega_{\mathrm{c}}}
\sum_{p=1}^{\ell}
\left(
e^{-\beta_{2p-1}\omega_{\mathrm{c}}} 
- 
e^{-\beta_{2p}\omega_{\mathrm{c}}}
\right) 
\le 0. 
\end{align*}
Then, Theorem 1.3 of Ref.\cite{hir99} says that 
the ground-state energy of the Rabi Hamiltonian is expressed as 
\begin{equation}
E_{\mathrm{Rabi}}
=
\frac{\hbar\omega_{\mathrm{c}}}{2} 
-\, \frac{\hbar\mathrm{g}^{2}}{\omega_{\mathrm{c}}} 
-\, \lim_{\beta\to\infty}\frac{\hbar}{\beta}
\ln
\left\{
I_{\mathrm{even}}(\beta)+I_{\mathrm{odd}}(\beta)
\right\}
\label{eq:gse-sb1}
\end{equation}
for arbitrary coupling constant $\mathrm{g}$ 
provided that $1/2\le\omega_{\mathrm{c}}/\omega_{\mathrm{a}}$.

Eq.(\ref{eq:gse-sb1}) is strict, 
but the expression is very complicated 
because those of the functions 
$I_{\mathrm{even}}(\beta)$ and $I_{\mathrm{odd}}(\beta)$ are so. 
Thus, we can make it simpler with a constant. 
We modify the functions 
$I_{\mathrm{even}}(\beta)$ and $I_{\mathrm{odd}}(\beta)$ by replacing 
the constants $G_{\beta_{1},\cdots,\beta_{2\ell}}$ 
and $F_{\beta_{1},\cdots,\beta_{2\ell+1}}$ in them 
with simple constants $\ell G$ and $G/2$, respectively, 
for an arbitrary parameter $G$ in the closed interval 
$[-1 , 0]$: 
\begin{align*}
& I_{\mathrm{even}}^{G}(\beta):= 
\mathrm{cosh}
\left[
(\beta\omega_{\mathrm{a}}/2)
e^{-2\mathrm{g}^{2}(G+1)/\omega_{\mathrm{c}}^{2}}
\right], \\ 
& I_{\mathrm{odd}}^{G}(\beta):= 
\mathrm{sinh}
\left[
(\beta\omega_{\mathrm{a}}/2)
e^{-2\mathrm{g}^{2}(G+1)/\omega_{\mathrm{c}}^{2}}
\right].
\end{align*} 
Then, Theorem 1.5 of Ref.\cite{hir99} 
says that for every coupling constant $\mathrm{g}$ 
we can uniquely determine a constant $G(\mathrm{g})$  
in the closed interval $[-1 , 0]$
so that the ground-state energy $E_{\mathrm{Rabi}}$ 
turns out to be a simple expression: 
\begin{align}
E_{\mathrm{Rabi}} 
&=
\frac{\hbar\omega_{\mathrm{c}}}{2}  
-\, \frac{\hbar\mathrm{g}^{2}}{\omega_{\mathrm{c}}}
- \lim_{\beta\to\infty}
\frac{\hbar}{\beta}
\ln
\left\{
I_{\mathrm{even}}^{G(\mathrm{g})}(\beta)
+ I_{\mathrm{odd}}^{G(\mathrm{g})}(\beta)
\right\}
\notag \\ 
&=
\frac{\hbar\omega_{\mathrm{c}}}{2} 
-\, \frac{\hbar\mathrm{g}^{2}}{\omega_{\mathrm{c}}}
-\, \frac{\hbar\omega_{\mathrm{a}}}{2}
\mathrm{exp}
\left[
- 2\frac{\mathrm{g}^{2}}{\omega_{\mathrm{c}}^{2}}
\left( G(\mathrm{g})+1\right) 
\right].   
\label{eq:gse-sb2}
\end{align}
The constant $G(\mathrm{g})$ is 
determined as a solution of the equation: 
$$
\lim_{\beta\to\infty}
\left\{
\frac{
I_{\mathrm{even}}(\beta)-I_{\mathrm{odd}}(\beta))}
{I_{\mathrm{even}}^{G(\mathrm{g})}(\beta)-I_{\mathrm{odd}}^{G(\mathrm{g})}(\beta)}
\right\}^{1/\beta}=1.
$$ 
In fact, we have the limit $G(\mathrm{g})\to 0$ as 
$\mathrm{g}\to\infty$. 
The factor $2\hbar\mathrm{g}^{2}(G(\mathrm{g})+1)/\omega_{\mathrm{c}}^{2}$ 
in Eq.(\ref{eq:gse-sb2}) plays a role similar to the classical action 
associated with a single-instanton solution (see Eq.(3.36) of Ref.\cite{as10}) 
in the expression of the transition amplitude. 
The functions $I_{\mathrm{even}}^{G(\mathrm{g})}(\beta)$ and 
$I_{\mathrm{odd}}^{G(\mathrm{g})}(\beta)$ 
correspond to Eqs.(3.41) of Ref.\cite{as10}.  
By taking $-1$ and $0$ as the constant $G(\mathrm{g})$ 
in Eq.(\ref{eq:gse-sb2}), 
we obtain estimates (\ref{eq:estimate2}) 
as the roughest estimates derived from Eq.(\ref{eq:gse-sb2}).

\section{Proofs of Some Mathematical Statements} 
\label{sec:proofs} 

\subsection{Proof of (\ref{eq:JCGSE-estimates})}
\label{subsec:proof-JCGSE-estimates}

In this subsection we prove the estimates (\ref{eq:JCGSE-estimates}). 
Define the parameterized asymptotic Hamiltonian $H_{\mathrm{asym}}(\varepsilon)$ 
in the sense described in \S\ref{sec:SUSY-Breaking} by 
$$
H_{\mathrm{asym}}(\varepsilon):=
\hbar\omega_{\mathrm{c}}(\varepsilon)
\left( a^{\dagger}a+\frac{1}{2}
\right)
+\hbar\mathrm{g}
\sigma_{x}\left( a^{\dagger}+a\right).
$$
Then, this is decomposed as: 
\begin{equation}
H_{\mathrm{asym}}(\varepsilon)
=\frac{1}{2}
\left( H_{\mathrm{JC}}^{\mathrm{g}}(\varepsilon)
+\sigma_{x}H_{\mathrm{JC}}^{\mathrm{g}}(\varepsilon)\sigma_{x}
\right).  
\label{eq:decomposition-asym-Hamiltonian}
\end{equation}
It is easy to get the inequality: 
\begin{eqnarray*}
E_{\mathrm{JC}}^{\mathrm{g}}(\varepsilon)\|\psi\|^{2}
&\le&  
\frac{1}{2}
\left\{ 
\langle\psi |H_{\mathrm{JC}}^{\mathrm{g}}(\varepsilon)|\psi\rangle
+\langle\psi |\sigma_{x}H_{\mathrm{JC}}^{\mathrm{g}}(\varepsilon)\sigma_{x}|
\psi\rangle
\right\} \\ 
&=& 
\langle\psi |H_{\mathrm{asym}}(\varepsilon)|\psi\rangle, 
\end{eqnarray*}
where $\|\quad\|$ is the norm induced 
by the inner product: $\|\phi\|:=
\langle\phi |\phi\rangle^{1/2}$.  
This inequality tells us that the ground-state energy of 
$H_{\mathrm{JC}}^{\mathrm{g}}(\varepsilon)$ does not exceed that of 
the parameterized asymptotic Hamiltonian 
$H_{\mathrm{asym}}(\varepsilon)$: 
$$
E_{\mathrm{JC}}^{\mathrm{g}}(\varepsilon)
\le 
\inf\mathrm{Spec}(H_{\mathrm{asym}}(\varepsilon)).  
$$
Using the unitary operator $U_{0}$ defined by 
\begin{equation}
U_{0}:=\frac{1}{\sqrt{2}}(I-i\sigma_{y})
=\frac{1}{\sqrt{2}}
\begin{pmatrix}
1 & -1 \\ 
1 & 1
\end{pmatrix}, 
\label{eq:U0}
\end{equation} 
we have 
\begin{align}
U_{0}^{*}H_{\mathrm{asym}}(\varepsilon)U_{0}
=&
\frac{ I+\sigma_{z}}{2}\otimes
\left\{
\hbar\omega_{\mathrm{a}}(\varepsilon)
\left( a^{\dagger}a+\frac{1}{2}
\right)
+\hbar\mathrm{g}
\left( a^{\dagger}+a\right)
\right\}
\notag \\ 
& 
+
\frac{ I-\sigma_{z}}{2}\otimes
\left\{
\hbar\omega_{\mathrm{a}}(\varepsilon)
\left( a^{\dagger}a+\frac{1}{2}
\right)
-\hbar\mathrm{g}
\left( a^{\dagger}+a\right)
\right\}.  
\label{eq:trans-asym-Hamiltonian} 
\end{align}
Using Eqs. (\ref{eq:bogoliubov}) 
and (\ref{eq:trans-asym-Hamiltonian}), 
the ground-state energy of $H_{\mathrm{asym}}(\varepsilon)$ is 
expressed as: 
$$
\inf\mathrm{Spec}(H_{\mathrm{asym}}(\varepsilon))=\, 
-\, \frac{\hbar\mathrm{g}^{2}}{\omega_{\mathrm{c}}(\varepsilon)}.
$$
Thus, we have 
our desired upper bound of 
$E_{\mathrm{JC}}^{\mathrm{g}}(\varepsilon)$. 

We note the equation:
\begin{equation}
\| a\psi\|
=
\langle a\psi | a\psi\rangle^{1/2}
= 
\langle\psi |a^{\dagger}a|\psi\rangle^{1/2} 
\label{eq:equation-annihilation}
\end{equation}
for every state $\psi$. 
Noting the canonical commutation relation 
(CCR), $[a,a^{\dagger}]:=aa^{\dagger}-a^{\dagger}a=1$, 
we have the following inequality 
in the same way as above: 
\begin{align}
\| a^{\dagger}\psi\|
&=
\langle\psi |aa^{\dagger}|\psi\rangle^{1/2} 
\notag \\ 
&= 
\left\{  
\|\psi\|^{2}
+
\langle\psi |a^{\dagger}a|\psi\rangle
\right\}^{1/2}
\le 
\|\psi\|
+ 
\langle\psi |a^{\dagger}a|\psi\rangle^{1/2}. 
\label{eq:inequality-creation}
\end{align}
Using the Schwarz inequality, 
the estimate by the operator norm $\|\quad\|_{\mathrm{op}}$, 
Eq.(\ref{eq:equation-annihilation}), and the inequality 
(\ref{eq:inequality-creation}), 
we have 
\begin{align*}
|\langle\psi |W_{\mathrm{R}}|\psi\rangle| 
\le& 
\|\psi\|
\bigl\{
\|\sigma_{-}\|_{\mathrm{op}}
\| a^{\dagger}\psi
\|_{\mathcal{H}}  
+
\|\sigma_{+}\|_{\mathrm{op}}
\| a\psi
\|_{\mathcal{H}} 
\bigr\} \\ 
\le& 
\|\psi\|_{\mathcal{H}}^{2}
+ 
2\|\psi\|_{\mathcal{H}}
\langle\psi | a^{\dagger}a|\psi\rangle^{1/2}, 
\end{align*}
which implies 
\begin{align}
|\langle\psi |\hbar\mathrm{g}W_{\mathrm{R}}|\psi\rangle| 
&\le 
\hbar\mathrm{g}
\left\{
\delta
\langle\psi |a^{\dagger}a|\psi\rangle
+\left( 
1+\frac{1}{\delta}
\right)
\|\psi\|^{2}
\right\}
\notag  \\
&\le 
\frac{\delta\mathrm{g}}{\omega_{\mathrm{c}}(\varepsilon)}
\langle\psi |H_{0}(\varepsilon)|\psi\rangle 
+
\hbar\mathrm{g}
\left\{ 
\frac{\delta\omega_{\mathrm{a}}(\varepsilon)
}{2\omega_{\mathrm{c}}(\varepsilon)}
+
\left( 
1+\frac{1}{\delta}
\right)
\right\}
\|\psi\|^{2} 
\label{eq:estimate-WR} 
\end{align}
for every $\delta>0$. 
Here the parameterized free Hamiltonian $H_{0}(\varepsilon)$ 
was defined in Eq.(\ref{eq:parameterized-free-Hamiltonian}), 
and  we used the fact that 
$(\hbar\omega_{\mathrm{a}}(\varepsilon)/2)
\sigma_{z}\ge -\hbar\omega_{\mathrm{a}}(\varepsilon)/2$. 
The inequality (\ref{eq:estimate-WR}) leads to 
\begin{align*}
\langle\psi |H_{\mathrm{JC}}^{\mathrm{g}}(\varepsilon)|\psi\rangle 
=& 
\langle\psi |H_{0}(\varepsilon)|\psi\rangle
+ \langle\psi |\hbar\mathrm{g}W_{\mathrm{R}}\psi\rangle \\ 
\ge& 
\langle\psi |H_{0}(\varepsilon)|\psi\rangle 
-
|\langle\psi |\hbar\mathrm{g}W_{\mathrm{R}}|\psi\rangle|  \\ 
\ge&
\left(
1-\frac{\delta\mathrm{g}}{\omega_{\mathrm{c}}(\varepsilon)}
\right)
\langle\psi |H_{0}(\varepsilon)|\psi\rangle \\ 
&\qquad 
-\hbar\mathrm{g}\left\{ 
\frac{\delta\omega_{\mathrm{a}}\left(\varepsilon\right)
}{2\omega_{\mathrm{c}}\left(\varepsilon\right)}
+
\left( 1+\frac{1}{\delta}
\right)
\right\}
\|\psi\|^{2} 
\end{align*} 
Take $\delta$ as 
$\delta=\omega_{\mathrm{c}}(\varepsilon)/\mathrm{g}$ now. 
Then, we eventually reach the inequality: 
$$
-\,\frac{\hbar\omega_{\mathrm{a}}(\varepsilon)}{2} 
-\hbar\mathrm{g}
-\,\frac{\hbar\mathrm{g}^{2}}{\omega_{\mathrm{c}}(\varepsilon)}
\le 
\frac{
\langle\psi |H_{\mathrm{JC}}^{\mathrm{g}}(\varepsilon)|\psi\rangle 
}{\|\psi\|_{\mathcal{H}}^{2}}, 
$$ 
which implies our desired lower bound of 
$E_{\mathrm{JC}}^{\mathrm{g}}(\varepsilon)$. 

\subsection{Proof of (\ref{eq:math-statement1})} 
\label{subsec:proof-math-statement1}

We assume that the coupling strength $\mathrm{g}$ 
runs over the interval $I_{\varepsilon}$ 
defined in Eq.(\ref{eq:interval}) 
to prove the statement (\ref{eq:math-statement1}). 
So, the coupling strength $\mathrm{g}$ 
satisfies $\mathrm{g}^{2}\le G_{1}(\varepsilon)$. 
Here the quantity $G_{|\nu|}(\varepsilon)$ was 
defined in Eq.(\ref{eq:quantity-G}). 
Since we immediately have the equation, 
$G_{|\nu|+1}(\varepsilon)-
G_{|\nu|}(\varepsilon)=(1-\varepsilon)^{2}\omega^{2}>0$, 
we reach the inequality,  
\begin{equation}
G_{|\nu|}(\varepsilon)<G_{|\nu|+1}(\varepsilon),\qquad 
\nu=\, -1, -2, \cdots.
\label{eq:G<G}
\end{equation}
Thus, since the condition 
$\mathrm{g}^{2}<\varepsilon^{2}G_{|\nu|}(\varepsilon)$ always holds 
for every coupling strength $\mathrm{g}\in I_{\varepsilon}$ 
and each $\nu=-2, -3, \cdots$ 
by the inequality (\ref{eq:G<G}), 
we realize that  
\begin{equation}
\nu_{*}=0\quad 
\textrm{as long as $\mathrm{g}\in I_{\varepsilon}$}.
\label{eq:standard-g}
\end{equation} 
In addition, the mathematical fact (\ref{eq:criterion1}) 
and the inequality (\ref{eq:G<G}) 
say that $E_{0}^{\mathrm{g}}(\varepsilon)<
E_{-|\nu|}^{\mathrm{g}}(\varepsilon)$ 
for all $\nu=\, -1, -2, \cdots$. 

We pay our attention to the case $\varepsilon=0$ now. 
Then, the mathematical fact (\ref{eq:criterion1}) 
also says that there is a strictly negative 
integer $\nu$ so that 
$E_{0}^{\mathrm{g}}(0)>E_{-|\nu|}^{\mathrm{g}}(0)$ 
if and only if $\mathrm{g}^{2}>G_{1}(0)$. 
Thus, applying this fact to the chiral-counter Hamiltonian 
$H_{\mathrm{JC}}^{\mathrm{g}/\varepsilon}(0)$, 
the index $\nu_{**}$ of the GSTI satisfies
\begin{equation}
\begin{cases}
\nu_{**}=0\quad 
\textrm{if and only if 
$\mathrm{g}^{2}/\varepsilon^{2}\le G_{1}(0)=\omega^{2}$}, \\ 
\nu_{**}<0\quad 
\textrm{if and only if 
$\mathrm{g}^{2}/\varepsilon^{2}>G_{1}(0)=\omega^{2}$}.  
\end{cases}
\label{eq:chiral-g}
\end{equation} 
Meanwhile, we have 
\begin{align*}
G_{1}(\varepsilon)
&\equiv 
(1-\varepsilon)^{2}\omega^{2}
\left( 
1+\frac{\Delta_{\varepsilon}}{(1-\varepsilon)\omega}
\right)  
\ge 
(1-\varepsilon)^{2}\omega^{2}
\left( 
1+\frac{\Delta_{0}}{(1-\varepsilon)\omega}
\right) \\ 
&\ge 
(1-\varepsilon)^{2}\omega^{2}
\left( 
1+\frac{\Delta_{0}}{\omega}
\right)  
\ge 
\varepsilon^{2}\omega^{2}
\left( 
1+\frac{\Delta_{0}}{\omega}
\right) 
=\varepsilon^{2}G_{1}(0)
=\mathrm{g}[\varepsilon]^{2}.  
\end{align*} 
Here we used the inequality $\Delta_{\varepsilon}>\Delta_{0}=0$ 
at the first inequality, 
and the inequality $0<1-\varepsilon<1$ at the second inequality. 
Third inequality follows from 
$(1-\varepsilon)^{2}>\varepsilon^{2}$, 
caused by $0<\varepsilon<1/2$. 
Thus, the critical coupling constant $\mathrm{g}[\varepsilon]$ 
is in the interval $I_{\varepsilon}$. 
Accordingly, if the coupling strength $\mathrm{g}$ 
satisfies $\mathrm{g}<\mathrm{g}[\varepsilon]$, 
then it is also in the interval $I_{\varepsilon}$. 
We can conclude the proof of our desired 
statement (\ref{eq:math-statement1}) 
by combining Eqs.(\ref{eq:chiral-g}) and (\ref{eq:standard-g}). 
 
\subsection{Proof of (\ref{eq:estimate2})} 
\label{subsec:proof-estimate2}

We give the proof of the estimate (\ref{eq:estimate2}) 
in this subsection. 
To begin with, we recall the value of the following inner product 
for every real number $\gamma$ and the Fock vacuum $|0\rangle$: 
\begin{equation}
\langle 0|
e^{\pm\gamma\mathrm{g}(a^{\dagger}-a)/\omega_{\mathrm{c}}(\varepsilon)}
|0\rangle
=e^{-\gamma^{2}\mathrm{g}^{2}/(2\omega_{\mathrm{c}}(\varepsilon)^{2})}. 
\label{eq:0-expectation}
\end{equation}
We recall the equations, 
$-\sigma_{x}=U_{0}^{*}\sigma_{z}U_{0}$ 
and 
$\sigma_{z}=U_{0}^{*}\sigma_{x}U_{0}$, 
for the unitary operator $U_{0}$ defined in Eq.(\ref{eq:U0}), 
and the Pauli matrices $\sigma_{x}$ and $\sigma_{z}$. 
Thus, the LHS of the estimates (\ref{eq:estimate2}) 
follows from the simple variational principle 
with the matrix $U_{0}$: 
\begin{align}
e_{\mathrm{low}}(\mathrm{g}) 
&\le
-(\hbar\omega/2)\langle U_{0}^{*}\psi |\sigma_{x}
|U_{0}^{*}\psi\rangle 
\notag \\ 
&\qquad 
+
\langle U_{0}^{*}\psi |
\left\{
     \hbar\omega\left( a^{\dagger}a+1/2\right)
        +\hbar\mathrm{g}\sigma_{z}\otimes (a^{\dagger}+a)
  \right\}
|U_{0}^{*}\psi\rangle 
\notag \\ 
&= \langle\psi |H_{\mathrm{R}}|\psi\rangle. 
\label{eq:expression-HR}
\end{align} 
Here we used Eq.(\ref{eq:bogoliubov}) to estimate 
the second term of the middle expression 
from below.

To derive the RHS of the estimates (\ref{eq:estimate2}), 
we insert a special vector $\phi$ given by 
$\phi:=U_{0}\phi_{0}$ 
into the vector $\psi$ in the expression (\ref{eq:expression-HR}) 
of the inner product $\langle\psi |H_{\mathrm{R}}|\psi\rangle$, 
where $\phi_{0}$ is defined by 
$$
\phi_{0}:=
\frac{1}{\sqrt{2}}
\left\{ 
|\!\!\uparrow\rangle
e^{-\mathrm{g}(a^{\dagger}-a)/\omega}|0\rangle 
+
|\!\!\downarrow\rangle
e^{+\mathrm{g}(a^{\dagger}-a)/\omega}|0\rangle
\right\}.
$$  
Then, we have the upper bound 
$e_{\mathrm{upp}}(\mathrm{g})$: 
\begin{align*}
E_{\mathrm{Rabi}} 
\le& 
\langle\phi |H_{\mathrm{Rabi}}|\phi\rangle \\ 
=&
-(\hbar\omega/2)\langle 0|\sigma_{x}|0\rangle 
+
\langle 0|\left\{
\hbar\omega
\left( a^{\dagger}a+1/2\right)
+\hbar\mathrm{g}\sigma_{z}\otimes (a^{\dagger}+a)
\right\}
|0\rangle \\ 
=& 
e_{\mathrm{upp}}(\mathrm{g}).
\end{align*} 
Here we respectively used (\ref{eq:0-expectation}) and 
Eq.(\ref{eq:bogoliubov}) to compute the first term 
and the second term of the middle expression. 

\subsection{Proofs of [I]--[III] and (\ref{eq:GSTI-relation})} 
\label{subsec:proof-I-III}

First, we give expressions of 
the lower bond $E_{\mathrm{lbd}}^{\mathrm{g}}(\varepsilon)$ 
and the upper bound $E_{\mathrm{ubd}}^{\mathrm{g}}(\varepsilon)$ 
defined in Eqs.(\ref{eq:lu-bound}). 
The direct computation using the Eqs.(\ref{eq:eigenenergy-n}) 
gives concrete expressions of 
the lower bound $E_{\mathrm{lbd}}^{\mathrm{g}}(\varepsilon)$ 
and the upper bound $E_{\mathrm{ubd}}^{\mathrm{g}}(\varepsilon)$:

For any GSTI $[\, |\nu_{*}|,|\nu_{**}|\, ](\varepsilon)$ 
with $\nu_{*}=0$ and $\nu_{**}=\, -1, -2, \cdots$, 
we can compute the both bounds as:
\begin{equation}
\begin{cases}
{\displaystyle 
E_{\mathrm{lbd}}^{\mathrm{g}}(\varepsilon)
=\hbar\omega 
\Bigl[
-\varepsilon(|\nu_{**}|-1)
+\frac{\mathrm{g}}{\omega}\sqrt{|\nu_{**}|}
-\frac{\mathrm{g}^{2}}{\omega^{2}}
\Bigr]
},  \\ 
\qquad \\ 
{\displaystyle 
E_{\mathrm{ubd}}^{\mathrm{g}}(\varepsilon)  
= 
\hbar\omega 
\Bigl[
\frac{1}{2}-\varepsilon(|\nu_{**}|-1)
+\frac{\mathrm{g}}{\omega}\sqrt{|\nu_{**}|}
-\frac{\mathrm{g}^{2}}{\omega^{2}}
-\frac{1}{2}e^{-2\mathrm{g}^{2}/\omega^{2}}
\Bigr]
}. 
\end{cases}
\label{eq:diff-estimates-1'}
\end{equation}

For any GSTI $[\, |\nu_{*}|,|\nu_{**}|\, ](\varepsilon)$ 
with $\nu_{*}, \nu_{**}=\, -1, -2, \cdots$, 
the lower and upper bounds are respectively expressed as: 
\begin{equation}
\begin{cases}
{\displaystyle 
E_{\mathrm{lbd}}^{\mathrm{g}}(\varepsilon) 
= 
\hbar\omega
\Bigl[
-|\nu_{*}|
-\varepsilon(|\nu_{**}|-|\nu_{*}|)
+\sqrt{
\varepsilon^{2}
+\frac{\mathrm{g}^{2}}{\omega^{2}}
|\nu_{*}|}}  \\ 
\qquad\qquad\qquad\qquad
{\displaystyle  
+\frac{\mathrm{g}}{\omega}
\sqrt{|\nu_{**}|}
-\, \frac{\mathrm{g}^{2}}{\omega^{2}}
\Bigr]},  \\ 
\qquad \\ 
{\displaystyle 
E_{\mathrm{ubd}}^{\mathrm{g}}(\varepsilon)  
= 
\hbar\omega
\Bigl[
\frac{1}{2}-|\nu_{*}|
-\varepsilon(|\nu_{**}|-|\nu_{*}|) 
+\sqrt{
\varepsilon^{2}
+\frac{\mathrm{g}^{2}}{\omega^{2}}
|\nu_{*}|}}  \\ 
\qquad\qquad\qquad\qquad 
{\displaystyle  
+\frac{\mathrm{g}}{\omega}
\sqrt{|\nu_{**}|}
-\, \frac{\mathrm{g}^{2}}{\omega^{2}}
-\, \frac{1}{2}e^{-2\mathrm{g}^{2}/\omega^{2}}
\Bigr]}. 
\end{cases}
\label{eq:diff-estimate-2} 
\end{equation}

\textit{Proof of [I]}: 
We consider the weak and strong coupling regimes 
given by $0\le\mathrm{g}/\omega\le\varepsilon<0.1$ now. 
In these regimes, the coupling strength $\mathrm{g}$ 
satisfies the condition $\mathrm{g}\le\mathrm{g}[\varepsilon]
=\varepsilon\omega$. 
So, the statement (\ref{eq:math-statement1}) says 
that the GSTI are $[0,0](\varepsilon)$. 
Thus, the estimates (\ref{eq:lu-estimate-ws}) 
follow from Eqs.(\ref{eq:diff-estimates-1'}) and 
the inequality (\ref{eq:estimate1}). 

\begin{figure}[htbp]
  \begin{center}
  \resizebox{80mm}{!}{\includegraphics{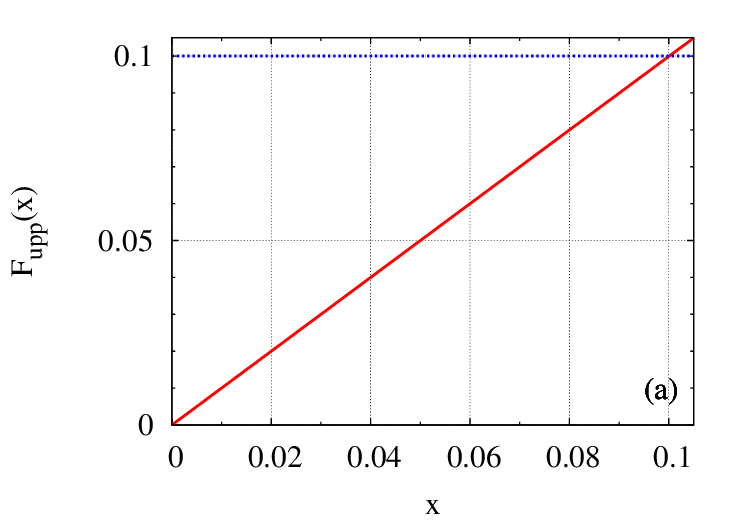}}
\qquad 
  \resizebox{80mm}{!}{\includegraphics{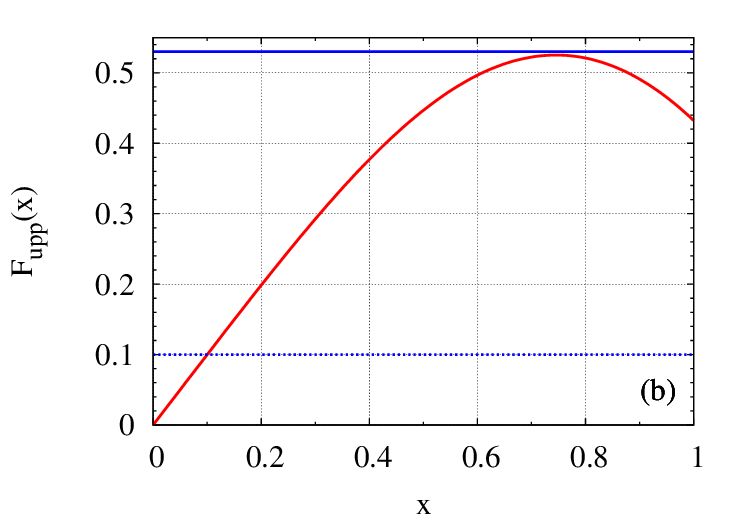}}   
  \end{center}
\vspace{3mm}
  \caption{\scriptsize 
$F_{\mathrm{upp}}(x)$ (a) $0\le x\le 0.1$. 
(b) $0\le x\le 1$.}
\label{fig:F_upp} 
\end{figure}

Define the function $F_{\mathrm{upp}}(x)$ by 
\begin{equation}
F_{\mathrm{upp}}(x):= 
\frac{1}{2}+x-x^{2}-\frac{1}{2}e^{-2x^{2}},
\qquad x\ge 0, 
\label{eq:F_upp}
\end{equation}
and take the decomposition rate $\varepsilon$ as 
$\varepsilon=\mathrm{g}/\omega$. 
Then, the upper bound in the estimate 
(\ref{eq:lu-estimate-ws}) is 
$\hbar\omega F_{\mathrm{upp}}(\mathrm{g}/\omega)$. 
Meanwhile, $0\le F_{\mathrm{upp}}(x)\le 0.1$ 
for every $x$ with $0\le x\le 0.1$ 
(See Fig.\ref{fig:F_upp}(a)). 
Thus, we obtain the upper bound (\ref{eq:estimate-0.0}).   

\textit{Proof of [II]}: 
Let us take the GSTI as $[0 , 1](\varepsilon)$ now. 
The estimate (\ref{eq:lu-estimate-u1}) follows 
directly from 
Eqs.(\ref{eq:diff-estimates-1'}) and 
the inequality (\ref{eq:estimate1}). 
The upper bound in the estimate 
(\ref{eq:lu-estimate-u1}) is 
$\hbar\omega F_{\mathrm{upp}}(\mathrm{g}/\omega)$. 
It is clear that $0\le F_{\mathrm{upp}}(x)\le 1/2$ 
for every $x$ with $1<x$. 
Meanwhile, we can show that 
$0\le F_{\mathrm{upp}}(x)\le 0.53$ 
for every $x$ with $0\le x\le 1$ 
(See Fig.\ref{fig:F_upp}(b)). 
Thus, we reach the upper bound (\ref{eq:estimate-0.1}).   

\textit{Proof of [III]}: 
Let us take the GSTI as $[\, |\nu_{*}|\, ,\, |\nu_{**}|\,](\varepsilon)$ 
with $0<|\nu_{*}|\le |\nu_{**}|+1$ now. 
Then, the estimate (\ref{eq:lu-estimate-u2}) follows 
directly from Eqs.(\ref{eq:diff-estimate-2}) and 
the inequality (\ref{eq:estimate1}). 
Using the fact that $\sqrt{a^{2}+b^{2}}\le a+b$ 
for every non-negative number $a$ and $b$, 
we can bound the upper bound 
$E_{\mathrm{ubd}}^{\mathrm{g}}(\varepsilon)$ as: 
\begin{align}
E_{\mathrm{ubd}}^{\mathrm{g}}(\varepsilon) 
&\le 
\hbar\omega
\Bigl[
\frac{1}{2}-|\nu_{*}|
-\varepsilon\left(|\nu_{**}|-|\nu_{*}|+1\right)
+2\sqrt{|\nu_{*}|}\, \frac{\mathrm{g}}{\omega}\, 
-\, \frac{\mathrm{g}^{2}}{\omega^{2}}
\Bigr] 
\notag \\ 
&\le 
\hbar\omega
\Bigl[ 
\frac{1}{2}-\varepsilon\left(|\nu_{**}|-|\nu_{*}|+1\right)
\Bigr]. 
\label{eq:new-bound-E_ubd}
\end{align} 
Consequently, we obtain the upper bound 
(\ref{eq:estimate-*.**}) 
provided that $|\nu_{*}|\le |\nu_{**}|+1$. 

\textit{Proof of (\ref{eq:GSTI-relation})}: 
We have the inequality between 
the ground-state energy $E_{\mathrm{Rabi}}$ 
and the lowest-energy sum $E_{\mathrm{les}}(\varepsilon)$ 
given in (\ref{eq:les}) as  
\begin{align*}
E_{\mathrm{Rabi}}
&=
\langle\psi_{\mathrm{Rabi}}|H_{\mathrm{Rabi}}|\psi_{\mathrm{Rabi}}\rangle \\ 
&=
\langle\psi_{\mathrm{Rabi}}|
H_{\mathrm{JC}}^{\mathrm{g}}(\varepsilon)|
\psi_{\mathrm{Rabi}}\rangle 
+\varepsilon
\langle\psi_{\mathrm{Rabi}}|
H_{\mathrm{JC}}^{\mathrm{g}/\varepsilon}(0)|
\psi_{\mathrm{Rabi}}\rangle \\ 
&\ge 
E_{\mathrm{JC}}^{\mathrm{g}}(\varepsilon)
+\varepsilon E_{\mathrm{JC}}^{\mathrm{g}/\varepsilon}(0)
\equiv 
E_{\mathrm{les}}(\varepsilon). 
\end{align*}
Here we used the decomposition (\ref{eq:chiral-decomposition}). 
So, we realize by this inequality that 
the non-commutative energy 
$E_{\mathrm{diff}}(\varepsilon)$ is non-negative 
because of its definition, 
$E_{\mathrm{diff}}(\varepsilon)\equiv 
E_{\mathrm{Rabi}}-E_{\mathrm{les}}(\varepsilon)$. 
Thus, combining the inequalities (\ref{eq:estimate1}) and 
(\ref{eq:new-bound-E_ubd}), 
we reach the inequality: 
$$
0\le E_{\mathrm{diff}}(\varepsilon)
\le E_{\mathrm{ubd}}^{\mathrm{g}}(\varepsilon)
\le 
\hbar\omega
\Bigl[ 
\frac{1}{2}-\varepsilon\left(|\nu_{**}|-|\nu_{*}|+1\right)
\Bigr], 
$$
which implies (\ref{eq:GSTI-relation}).

\subsection{Proofs of (\ref{eq:gse-expansion}) 
and (\ref{eq:A-B-estimate})}
\label{sec:proof-A-B-estimate}

We define the subspace $\mathcal{H}_{+}$ by 
the space consisting of all superpositions of 
the states $|n,\uparrow\rangle$ and 
$|n+1,\downarrow\rangle$ for all even numbers $n$. 
Similarly, we give the subspace $\mathcal{H}_{-}$ by 
the space consisting of all superpositions of 
the states $|0,\downarrow\rangle$, $|n,\uparrow\rangle$, 
and $|n+1,\downarrow\rangle$ for all odd numbers $n$. 
Then, we realize that 
$\Pi\psi=(\sharp 1)\psi$ 
for wave functions $\psi\in\mathcal{H}_{\sharp}$, 
where $\Pi$ was the parity operator 
$\sigma_{z}(-1)^{a^{\dagger}a}$. 
We know that the ground state of the Rabi Hamiltonian 
is continuous with respect to the coupling strength $\mathrm{g}$. 
For instance, we can see it using the representation $H_{\mathrm{Rabi}}
=H_{\mathrm{asym}}+(\hbar\omega/2)\sigma_{z}$ 
with the expression (\ref{eq:parity}), 
together with the facts that 
the asymptotic Hamiltonian $H_{\mathrm{asym}}$ 
is solvable and that the ground state of 
the Rabi Hamiltonian is unique for every 
coupling strength $\mathrm{g}$ \cite{HH12}. 
In addition to this continuity, 
the Rabi Hamiltonian has the parity symmetry Eq.(\ref{eq:parity-symmetry}). 
So, the normalized ground state $\varphi_{\mathrm{Rabi}}$ belongs 
to the subspace $\mathcal{H}_{-}$: 

(i) $\varphi_{\mathrm{Rabi}}\in \mathcal{H}_{-}$, 
\hfill\break 
since it belongs to that subspace as $\mathrm{g}=0$: 
$\varphi_{\mathrm{Rabi}}(\mathrm{g}=0)=|0,\downarrow\rangle 
\in\mathcal{H}_{-}$. 

Meanwhile, the eigenstates 
$\{\varphi_{\nu}^{\mathrm{g}}(\varepsilon)\}_{\nu\in\mathbb{Z}}$ 
(resp. $\{\varphi_{\nu}^{\mathrm{g}/\varepsilon}(0)\}_{\nu\in\mathbb{Z}}$) 
of the parameterized JC Hamiltonian 
$H_{\mathrm{JC}}^{\mathrm{g}}(\varepsilon)$ 
(resp. $H_{\mathrm{JC}}^{\mathrm{g}/\varepsilon}(0)$) 
makes a complete orthonormal basis of our state space. 
Moreover, we have the relations: 
 
(ii) $\varphi_{0}^{\mathrm{g}}(\varepsilon), 
\varphi_{0}^{\mathrm{g}/\varepsilon}(0)\in\mathcal{H}_{-}$; 

(iii) $\varphi_{\pm|\nu|}^{\mathrm{g}}(\varepsilon), 
\varphi_{\pm|\nu|}^{\mathrm{g}/\varepsilon}(0)\in\mathcal{H}_{-}$ 
if $n=|\nu|-1$ is odd (i.e., $|\nu|$ is even); 

(iv) $\varphi_{\pm|\nu|}^{\mathrm{g}}(\varepsilon), 
\varphi_{\pm|\nu|}^{\mathrm{g}/\varepsilon}(0)\in\mathcal{H}_{+}$ 
if $n=|\nu|-1$ is even (i.e., $|\nu|$ is odd).    

The ground state of the Rabi Hamiltonian is 
expressed using its normalized ground state 
$\varphi_{\mathrm{Rabi}}$ as 
$E_{\mathrm{Rabi}}=\langle\varphi_{\mathrm{Rabi}}|H_{\mathrm{Rabi}}
|\varphi_{\mathrm{Rabi}}\rangle$. 
Applying the chiral decomposition (\ref{eq:chiral-decomposition}) 
to this, we have the equation
$$
E_{\mathrm{Rabi}}=
\langle\varphi_{\mathrm{Rabi}}|H_{\mathrm{JC}}^{\mathrm{g}}(\varepsilon)
|\varphi_{\mathrm{Rabi}}\rangle
+\varepsilon\langle\sigma_{x}\varphi_{\mathrm{Rabi}}
|H_{\mathrm{JC}}^{\mathrm{g}/\varepsilon}(0)
|\sigma_{x}\varphi_{\mathrm{Rabi}}\rangle. 
$$ 
Inserting the expansion of $\varphi_{\mathrm{Rabi}}$ 
by $\varphi_{\nu}^{\mathrm{g}}(\varepsilon)$ 
and the expansion of $\sigma_{x}\varphi_{\mathrm{Rabi}}$ 
by $\varphi_{\nu}^{\mathrm{g}/\varepsilon}(0)$, respectively, 
into the LHS of the individual inner products, 
we reach the expansion, 
$$
E_{\mathrm{Rabi}}=
\sum_{\nu\in\mathbb{Z}}
\left[ 
E_{\nu}^{\mathrm{g}}(\varepsilon)
|\langle\varphi_{\nu}^{\mathrm{g}}(\varepsilon)|
\varphi_{\mathrm{Rabi}}\rangle|^{2}
+
\varepsilon 
E_{\nu}^{\mathrm{g}/\varepsilon}(0)
|\langle\varphi_{\nu}^{\mathrm{g}/\varepsilon}(0)|
\sigma_{x}\varphi_{\mathrm{Rabi}}\rangle|^{2}
\right]. 
$$  
We can derive Eq.(\ref{eq:gse-expansion}) from this 
because the above properties (i)--(iv) 
concerning parity-symmetry 
and the fact $\sigma_{x}\varphi_{\mathrm{Rabi}}\in\mathcal{H}_{+}$ 
imply 
\begin{align}
& A_{\nu}=\langle\varphi_{\nu}^{\mathrm{g}}(\varepsilon)|
\varphi_{\mathrm{Rabi}}\rangle=0\,\,\,\,\, 
\text{if $|\nu|$ is odd}, 
\notag \\ 
& B_{\nu}=\langle\varphi_{\nu}^{\mathrm{g}/\varepsilon}(0)|
\sigma_{x}\varphi_{\mathrm{Rabi}}\rangle=0\,\,\,\,\, 
\text{if $|\nu|$ is even}.  
\label{eq:B-parity-symmetry}
\end{align}

We define the orthogonal projection operator 
$P_{n}$ by $P_{n}:=|n\rangle\langle n|$. 
Then, we have the expression of the number operator 
$N:=a^{\dagger}a$ 
as $N=\sum_{n=0}^{\infty}nP_{n}=\sum_{n=1}^{\infty}nP_{n}$, 
and thus, we have the equation  $P_{0}+N
=P_{0}+\sum_{n=1}^{\infty}nP_{n}$, 
which implies the operator inequality 
$P_{0}+N\ge 
\sum_{n=0}^{\infty}P_{n}=1$.   
Thus, we reach the operator inequality:
\begin{equation}
P_{0}\ge 1-N.
\label{eq:appendix2-1}
\end{equation} 

In the same way as in Lemma 4.3 of Ref.\cite{ah97}, we have the 
inequality: 
\begin{equation}
\langle\varphi_{\mathrm{Rabi}}|N|\varphi_{\mathrm{Rabi}}\rangle 
\le \frac{\mathrm{g}^{2}}{\omega^{2}},
\label{eq:appendix2-2}
\end{equation} 
where $\varphi_{\mathrm{Rabi}}$ is the normalized 
ground state with the ground-state energy 
$E_{\mathrm{Rabi}}$ of the Rabi Hamiltonian. 
Precisely, this is proved as follows: 
Using the commutator 
$[H_{\mathrm{Rabi}} , a]\varphi_{\mathrm{Rabi}}=
(H_{\mathrm{Rabi}}-E_{\mathrm{Rabi}})a\varphi_{\mathrm{Rabi}}$ 
with $[H_{\mathrm{Rabi}} , a]=\, 
-\hbar\omega a
-\hbar\mathrm{g}\sigma_{x}$, 
we reach the so-called pull-through formula, 
$$
a\varphi_{\mathrm{Rabi}}=\,
- \hbar\mathrm{g}
(H_{\mathrm{Rabi}}-E_{\mathrm{Rabi}}+\hbar\omega)^{-1}
\sigma_{x}\varphi_{\mathrm{Rabi}}.
$$ 
Applying this pull-through formula to the term 
$\langle\varphi_{\mathrm{Rabi}}|N|\varphi_{\mathrm{Rabi}}\rangle$ 
and using the Schwarz inequality, 
we obtain the inequality (\ref{eq:appendix2-2}). 

The two inequalities (\ref{eq:appendix2-1}) and 
(\ref{eq:appendix2-2}) imply the lower bound: 
\begin{equation}
\langle\varphi_{\mathrm{Rabi}}|P_{0}|\varphi_{\mathrm{Rabi}}\rangle  
\ge 1-\frac{\mathrm{g}^{2}}{\omega^{2}}. 
\label{eq:appendix2-3}
\end{equation} 
Meanwhile, the Schwarz inequality brings the upper bound:
\begin{equation}
\langle\varphi_{\mathrm{Rabi}}|P_{0}|\varphi_{\mathrm{Rabi}}\rangle  
\le
\|\varphi_{\mathrm{Rabi}}\|\, 
\| P_{0}\varphi_{\mathrm{Rabi}}\| 
\le 
\|\varphi_{\mathrm{Rabi}}\|^{2}=1. 
\label{eq:appendix2-4}
\end{equation} 

Since the equation 
$|A_{0}|^{2}+|B_{0}|^{2}
=\langle\varphi_{\mathrm{Rabi}}|P_{0}|\varphi_{\mathrm{Rabi}}\rangle$
follows from the straightforward computation, 
we eventually obtain our desired estimate 
(\ref{eq:A-B-estimate}) 
by Eq.(\ref{eq:B-parity-symmetry}), 
the bounds (\ref{eq:appendix2-3}) 
and (\ref{eq:appendix2-4}).

\end{document}